\documentclass[twoside]{article}
 \pdfoutput=1
\usepackage[a4paper]{geometry}
 \usepackage[utf8]{inputenc}
\usepackage[T1]{fontenc} 
\usepackage{RR}
\usepackage{booktabs}

\usepackage[cmex10]{amsmath}
\usepackage{amssymb}
\usepackage{subfig}
\usepackage{graphicx}
\usepackage[abs]{overpic}

\usepackage{units}

\usepackage{tikz}
\usepackage{url}
\usepackage{pgfplots}

\usepackage{color}

\newcommand{\SC}[1]{{}}

\RRNo{7981}
\RRdate{May 2012}
\RRauthor{
Emmanuel Agullo
\thanks[sfn]{INRIA, Hiepacs Project, 350 cours de la Libération, 33400 Talence, France. Email: Surname.Name@Inria.fr}%
  \and
Bérenger Bramas 
\thanksref{sfn}
\and
 Olivier Coulaud \thanksref{sfn} 
\and 
Eric Darve\thanks{Mechanical Engineering Department, Institute for Computational and Mathematical Engineering, Stanford University, Durand 209, 496 Lomita Mall, 94305-3030 Stanford, CA, USA. Email: darve@stanford.edu}
\and Matthias Messner \thanksref{sfn} 
\and Toru Takahashi
\thanks{Department of Mechanical Science and Engineering, Nagoya University, Japan. Email: ttaka@nuem.nagoya-u.ac.jp}
    }

\authorhead{Agullo \& Bramas \& others}
\RRtitle{Pipeline de la méthode multipôle rapide sur un moteur d'exécution}
\RRetitle{Pipelining the Fast Multipole Method over a Runtime System}
\titlehead{Pipelining the Fast Multipole Method over a Runtime System}
\RRresume{
Les méthodes multipoles rapides (FMM) sont une opération fondamentale
pour la simulation de nombreux problèmes physiques. Leur mise en
\oe{}uvre haute performance requiert habituellement d'optimiser
attentivement l'algorithme à la fois pour la physique visée et la
matériel utilisé. Dans ce papier, nous proposons une nouvelle approche
qui atteint une performance élevée et portable. Note méthode consiste
à exprimer l'algorithme FMM comme un flot de tâches et d'employer un
moteur d'exécution, StarPU, afin de traiter les tâches sur les
différentes unités d'exécution. Nous concevons précisément le flot de
tâches, les opérateurs mathématiques, leur implémentations sur unité
centrale de traitement (CPU) et processeur graphique (GPU) ainsi que
les schémas d'ordonnancement. Nous calculons les potentiels et forces
pour des problèmes de 200 millions de particules en 48.7 secondes sur
une machine homogène SGI Altix UV 100 comportant 16O c\oe{}urs et de
38 millions de particules en 13.34 secondes sur une machine hétérogène
composée d'un processeur Intel Nehalem accéléré avec 3 GPUs Nvidia
Fermi M2090.
}
\RRabstract{
Fast Multipole Methods (FMM) are a fundamental operation for the
simulation of many physical problems. The high performance design of
such methods usually requires to carefully tune the algorithm for both
the targeted physics and the hardware. In this paper, we propose a new
approach that achieves high performance
across architectures. Our method consists of expressing the FMM
algorithm as a task flow and employing a state-of-the-art runtime
system, StarPU, in order to process the tasks on the different
processing units. We carefully design the task flow, the mathematical
operators, their Central Processing Unit (CPU) and Graphics Processing
Unit (GPU) implementations, as well as scheduling schemes. We compute potentials and forces of 200 million
particles in 48.7 seconds on a homogeneous 160 cores SGI Altix UV 100
and of 38 million particles in 13.34 seconds on a heterogeneous 12 cores Intel Nehalem
processor enhanced with 3 Nvidia M2090 Fermi GPUs. }
\RRmotcle{Méthodes multipôles rapides, processeur graphique, architectures hétérogènes, moteur d'exécution, pipeline}
\RRkeyword{Fast multipole methods, graphics processing unit, heterogeneous architectures, runtime system, pipeline}

\RRprojets{Hiepacs}
\RCBordeaux 

\begin{document}
\makeRR   
\tableofcontents
\newpage

\section{Introduction}
\label{sec:introduction}

Pair-wise particle interactions play an important role in many physical
problems. Examples are astrophysical simulations, molecular dynamics, the
boundary element method, radiosity in computer-graphics and dislocation
dynamics. In the last decades numerous algorithms have been developed in order
to reduce the quadratic complexity of a direct computation. The fast multipole
method (FMM), first presented in~\cite{greengard:87}, are probably the most
prominent one. The fact that they have linear complexity makes them candidates
of first choice for processing large-scale simulations of physical
problems~\cite{greengard97}.  Thus, the design of efficient FMM
implementations is crucial for the high performance computing (HPC)
community. They have been ported to multicore
processors~\cite{Greengard199063,Chandramowlishwaran2010a, Cruz2011,
  EricDarve2011}, graphical processing units (GPUs)
clusters~\cite{Yokota20092066, Gumerov2008, 42tflops, takahashi:2012,Hu:2011,
  Lashuki2009}.  Because these codes were tightly coupled with the targeted
architectures, they could achieve excellent performance, sometimes beyond
$10^{6}$ floating point operations per second $(\unitfrac{GFlop}{s})$. On the
other hand, porting a code from one architecture to another was a commitment
requiring very important (valuable) human resources. In this paper, we
consider an alternative approach for achieving high performance across
architectures. For that purpose, we turn the FMM algorithm into a task flow
and employ a runtime system in order to dispatch the tasks on the actual
computational units.

Processing HPC algorithms over runtime systems was successfully studied in the
case of dense linear algebra algorithms during the past five years~\cite{faster, LU, QR,Quintana-Orti, quintana-orti2008scheduling,tileqr,SDLA,BosilcaBDFHHKLLLLYD11} and
is now a common utility for related state-of-the-art libraries such as
Plasma~\cite{PLASMA}, Magma~\cite{MAGMA} and Flame~\cite{FLAME}. Dense linear algebra were
excellent candidates for pioneering this path. First, the related task graph
is very wide and therefore allows for many computational units to run
concurrently. Second, the relative regularity of the tasks makes them
particularly easy to schedule and achieve optimum performance~\cite{ipdps2011}.
On the contrary, pipelining the FMM over a runtime system is much more
challenging. Indeed, the pattern of an FMM task flow is much more irregular
and, hence, involving to handle. Moreover, the nature of the different
computational steps and the subsequent granularity of tasks add complexity for
ensuring an efficient scheduling. To tackle these challenges, we carefully
analyze the FMM task flow such that it can be efficiently pipelined, we define
fast mathematical operators and implement them on central processing units
(CPU) and graphics processing units (GPUs), and we construct empirical
performance models together with appropriate scheduling algorithms.

We propose an innovative methodology for designing HPC FMM codes. To
our knowledge, only two other (very recent and not yet published) attempts
have been made for processing the FMM over runtime systems. Ltaief and
Yokota~\cite{ltaeif-yokota-2012} have studied the feasibility of the
approach on a 16 cores homogeneous
machine. Bordage~\cite{bordage-2012} obtained preliminary results for
the Helmholtz kernel on heterogeneous architectures. In the present
work, we not only show that the FMM can be highly pipelined, but also obtain
performance numbers across architectures which are comparable to well
established and heavily tuned methods for specific
architectures~\cite{42tflops,Lashuki2009,Hu:2011,EricDarve2011}.

The paper is organized as follows. In Section~\ref{sec:fmm}, we briefly
introduce the FMM algorithm in use and present an improved M2L kernel. In
Section~\ref{sec:fmmruntime}, we introduce the StarPU~\cite{AugThiNamWac10CCPE} runtime system and
explain how to build a naive FMM task flow to be processed by the runtime
system. In Section~\ref{sec:homogeneous}, we carefully design an improved
task flow and present scheduling strategies for homogeneous architectures and
assess their impact on multicore platforms. Finally, we tackle heterogeneous
architectures in Section~\ref{sec:heterogeneous} before concluding in
Section~\ref{sec:conclusion}.


%
%


\section{Fast Multipole Method} 
\label{sec:fmm}
Pair-wise particle interactions can be modeled mathematically as
\begin{equation}
  \label{eq:pairwise_summation}
  f_i = \sum_{j=1}^N P(x_i, y_j) \, w_j \quad \text{for } i=1,\dots,M. 
\end{equation}
Here, pairs of particles, in the following denoted as sources and targets, are
represented by their spatial position $x,y\in\mathbb{R}^3$, respectively. The
interaction is governed by the kernel function $P(x,y)$. The above summation
can also be understood as matrix-vector product $\mathsf{f}=\mathsf{Pw}$,
where $\mathsf{P}$ is a dense matrix of size $M\times N$. Hence, if we assume
$M\sim N$, the cost grows like $\mathcal{O}(N^2)$. The FMM reduces the cost of
computing such summations to $\mathcal{O}(N)$. Its idea is to approximate the
kernel function by a low-rank representation whose error decays exponentially
if sources and targets are well separated (far-field). If sources and targets
are not well separated (near-field) the kernel function will be used in its
original form. These facts are exploited by hierarchically partitioning the
computational domain into an octree (in $\mathbb{R}^3$) an subsequently
identifying the near and the far-field. Finally, we end up with a data-sparse
approximation of the dense matrix $\mathsf{P}$.

\subsection{The black-box FMM algorithm}
\label{sec:black-box-fmm}
Many FMM algorithms are kernel specific, meaning, they require a distinct
analytic treatment for each kernel function. Our approach is adopted
from~\cite{Fong2009} and can deal with broad range of kernel
functions. Examples for the Laplace and the Stokes kernel and for various
multiquadric basis functions have been presented. Another kernel independent
FMM with a lower computational cost is presented in~\cite{Ying:2004}. However,
it is less general, since it works only for functions which are fundamental
solutions of second-order constant coefficient non-oscillatory elliptic
partial differential equations. Our approach has also been extended to the
oscillatory Helmholtz kernel in~\cite{Messner2012}.

In the paper at hand we present results for the Laplace kernel function
\begin{equation}
  \label{eq:kernelfunctions}
  P(x,y) = \frac{1}{|x-y|} \quad \text{and} \quad F(x,y) = \frac{x-y}{|x-y|^3}.
\end{equation}
Note, the second one can be written as $F(x,y) = \nabla_x P(x,y)$. And we use
a Chebyshev interpolation scheme to interpolate these kernel functions as
\begin{equation}
  \label{eq:cheby_interpol}
  \begin{aligned}
    P(x,y) &\sim \sum_{m=1}^{\ell} S_m(x) \sum_{n=1}^{\ell} P(\bar x_m, \bar
    y_n) S_n(y) \quad \text{and}\\
    F(x,y) &\sim \sum_{m=1}^{\ell} \nabla_x S_m(x) \sum_{n=1}^{\ell} P(\bar
    x_m, \bar y_n) S_n(y).
  \end{aligned}
\end{equation}
by shifting the gradient $\nabla_x$ from $F(x,y)$ to the interpolation
polynomial $S_m(x)$. In the following the interpolation polynomial is referred
to as P2M (particle-to-moment), M2M (moment-to-moment), L2L (local-to-local)
and L2P (local-to-particle) operator and the point-wise evaluated kernel
function as the M2L (moment-to-local) operator. Note, the P2M, M2M, M2L and
L2L operators are the same for Equation~\eqref{eq:cheby_interpol}, only the
L2P operator differs.

In the remainder of this section, we present technical details on how we
improved the black-box FMM. Note, they are not required for the understanding
of the main scope of the paper and readers not familiar with the FMM may proceed
to Section~\ref{sec:fmmruntime}.

\subsubsection{Efficient implementation of the P2M, M2M, L2L and L2P
  operators}
In $\mathbb{R}^3$ we use the tensor-product ansatz for the translation
operators
\begin{equation}
  \label{eq:tensor_interpol}
  S_m(x) = S_{m,1}(x_1) \otimes S_{m,2}(x_2) \otimes S_{m,3}(x_3).
\end{equation}
where each
\begin{equation}
  \label{eq:interpol_polyn}
  S_{m,i}(x_i) = \frac{1}{\ell} + \frac{2}{\ell} \sum_{n=1}^{\ell-1}
  T_n(\bar x_{m,i}) T_n(x_i),
\end{equation}
is a polynomial of order $\ell-1$ and $T_n(x)$ with $x\in[-1,1]$ are Chebyshev
polynomials of first kind and $\bar x$ are interpolation points chosen to be
Chebyshev roots.

The application of the M2P and P2L operator requires $\mathcal{O}(M\ell^4)$
floating point operations. However, by exploiting the outer-product-like
representation from Equation~\eqref{eq:interpol_polyn}, the cost can be
reduced to $\mathcal{O}(M\ell^3 + \ell^4)$. Moreover, due to the
tensor-product ansatz in Equation~\eqref{eq:tensor_interpol} the
matrix-vector-product can be reformulated as a sequence of smaller
matrix-matrix-products coupled with permutations (use of Blas3). In this way
the cost for applying the M2M and L2L operator can be reduced from
$\mathcal{O}(\ell^6)$ to $\mathcal{O}(\ell^4)$ only.

\subsubsection{Improved M2L operator}
Normally, in $\mathbb{R}^3$ the far-field is limited to the at most $27$
near-field interactions of the parent-cell, only. This leads to at most $189$
far-field interactions for one cell. Most kernel functions are homogeneous (if
we scale the distance between source $y$ and target $x$ by a factor of
$\alpha$ the resulting potential is scaled by $\alpha^n$, where $n$ is a
constant and depends on the kernel function). Hence, the M2L kernels of all
$316$ possible far-field interactions for all cells need to be computed only
once.

The authors of \cite{Fong2009} proposed to compress all possible M2L operators
$\mathsf{P}_t$ with $t=1,\dots,316$, each of size $\ell^3\times \ell^3$, via
two big singular value decompositions (SVD). Subsequent algebraic
transformations lead to an efficient representation as $\mathsf{P}_t \sim
\mathsf{UC}_t\mathsf{V}^\top$, where $\mathsf{C}_t$ is a matrix of size
$r\times r$ (low-rank obtained via the SVD based on a prescribed accuracy
$\varepsilon_{\text{SVD}}$). The matrices $\mathsf{U}$ and $\mathsf{V}$, both
of size $\ell^3 \times r$, are the same for all $t$, hence, their application
can be shifted to the M2M and L2L operator, respectively.

We propose another approach. First, by exploiting symmetries, we can represent
the $316$ M2L operators by permutations of $16$, only. Second, we apply an
individual SVD to each of the $16$ operators. In other words, instead of
computing $316$ matrix-vector products we compute $16$ matrix-matrix-products
coupled with permutations.

\begin{table}[htbp]
  \caption{Comparison of both M2L optimizations}
  \label{tab:cost_of_one_vs_many_svds}
  \centering
    \begin{tabular}{c | c c | c}
      \toprule
      $Acc$ & $r_{316}$ & weighted $r_{16}$ & cost$_{16}/$ cost$_{316}$\\ 
      \midrule
      3 &  19 & 4.6  & 0.69 \\
      5 &  67 & 11.2 & 0.62 \\
      7 & 150 & 22.2 & 0.67 \\
      \bottomrule
  \end{tabular}
\end{table}
We compare both approaches in Tab.~\ref{tab:cost_of_one_vs_many_svds}. The
first column shows three different accuracies of the method given by $Acc$,
i.e., $(\ell,\varepsilon_{\text{SVD}}) = (Acc, 10^{-Acc})$. Studies in
\cite{Messner2012} have shown that it can be useful to correlate $\ell$ and
$\varepsilon_{\text{SVD}}$. In other words, by solely reducing
$\varepsilon_{\text{SVD}}$ or increasing $\ell$, the method does not give more
accurate results. The accuracies $Acc=3,5,7$ lead to an relative error of
$\varepsilon_{L_2} \sim 10^{-5}, 10^{-7}, 10^{-9}$, respectively. The second
column in Tab.~\ref{tab:cost_of_one_vs_many_svds} shows the obtained low-rank
$r_{316}$ by the first approach and the third column the weighted low-rank
$r_{16}$ of our approach. The overall cost is given by $\text{cost}_{316} =
\mathcal{O}\left(316 \cdot r_{316}^2\right)$ and $\text{cost}_{16} =
\mathcal{O}\left(316 \cdot 2\ell^3r_{16}\right)$, respectively. Column four
shows that our approach is favorable in terms of
\begin{itemize}
\item reduced precomputation cost and memory requirement: $16$ small SVD,
  each of size $\ell^3\times\ell^3$ instead of $2$ big SVDs, both of size $316
  \, \ell^3\times\ell^3$
\item better compression $(\text{weighted } r_{16}\ll r_{316})$ reduces cost
  of applying the M2L operators
\item better cache reuse: $16$ matrix-matrix products instead of $316$
  matrix-vector products
\end{itemize}


\section{FMM over a runtime system}
\label{sec:fmmruntime}

A runtime system is a software component that aims at supporting the execution
of an algorithm written in a relatively high-level language.  Different
runtime systems were designed to support accelerator-based
platforms. StarSs~\cite{GPUSs09Europar} is an annotation-based language that
can execute programs either on CPUs, GPUs or Cell processors, respectively,
using SMPSs, GPUSs or CellSs. Initially designed for distributed memory
machines, DAGuE~\cite{Bosilca201237} has been extended to handle GPUs too. The Harmony runtime system proposes features
similar to StarPU~\cite{1383447}. Sequoia~\cite{SEQUOIA_SC06} statically maps
hierarchical applications on top of clusters of hybrid machines. Charm++ can
also support GPUs~\cite{2009ChaNGaGPU}.

StarPU was originally designed for handling heterogeneous platforms such
as multicore chips accelerated with GPUs. It is a natural candidate
if one aims at achieving performance across architectures.

\subsection{StarPU runtime system}
\label{sec:starpu}

StarPU was essentially designed to execute codes on heterogeneous platforms.
The algorithm (FMM in our case) to be processed by StarPU
is written in a high-level language, oblivious to the underneath platform. It
is expressed as a so-called \emph{task flow} consisting of a set of tasks with
dependencies among them. Task flows can conveniently be represented as 
\emph{directed acyclic graphs} (DAGs) where vertices represent individual 
tasks and edges
dependencies among them. A multi-device
implementation of the tasks, so-called \emph{codelet}, is then provided by the
programmer. It gathers the kernel implementations available to the different
devices (CPU core and GPUs in our case). The data on which the tasks operate
may need to be moved between the different computational units. StarPU ensures
the coherency of these data. For that, data are registered to the runtime,
which is accessing them not through their memory address anymore but through a
StarPU abstraction, the \emph{handle}, returned by registration. StarPU
transparently guarantees that a task that needs to access a piece of data will
be given a pointer to a valid data replicate. It will take care of the data
movements and therefore relieve programmers from the burden of explicit data
transfers. StarPU also detects the CPU core the closest to a GPU and
selects it to handle that GPU.

In the next section, we show how the FMM, usually viewed as a tree
algorithm, can be turned into a task flow suitable for execution over
a runtime system.

\subsection{FMM task flow}
\label{sec:fmmtaskflow}
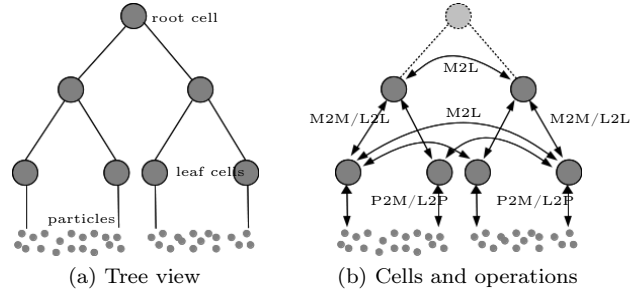
\begin{figure}[htbp]
  \centering
  \subfloat[Tree view]{
    \label{fig:unmirrored2}
    \begin{overpic}[scale=.5]{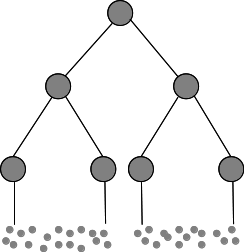}
      \put(52,88){\tiny root cell}
      \put(61,30){\tiny leaf cells}
      \put(13,12){\tiny particles}
    \end{overpic}
  }
  \hspace*{2em}
  \subfloat[Cells and operations]{
    \label{fig:unmirrored1}
    \begin{overpic}[scale=.5]{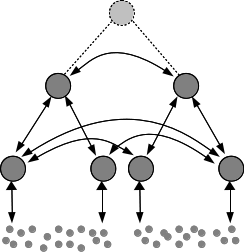}
      \put(13,19){\tiny P2M/L2P}
      \put(60,19){\tiny P2M/L2P}
      \put(-10,50){\tiny M2M/L2L}
      \put(80,50){\tiny M2M/L2L}
      \put(40,69){\tiny M2L}
      \put(41,52){\tiny M2L}
    \end{overpic}
  }
  \caption{Simplified FMM tree and subsequent operations.}
  \label{fig:unmirrored}
\end{figure}

\begin{figure}[htbp]
  \centering
  \subfloat[Mirrored tree view]{
    \label{fig:mirrored}
    \begin{overpic}[scale=.32]{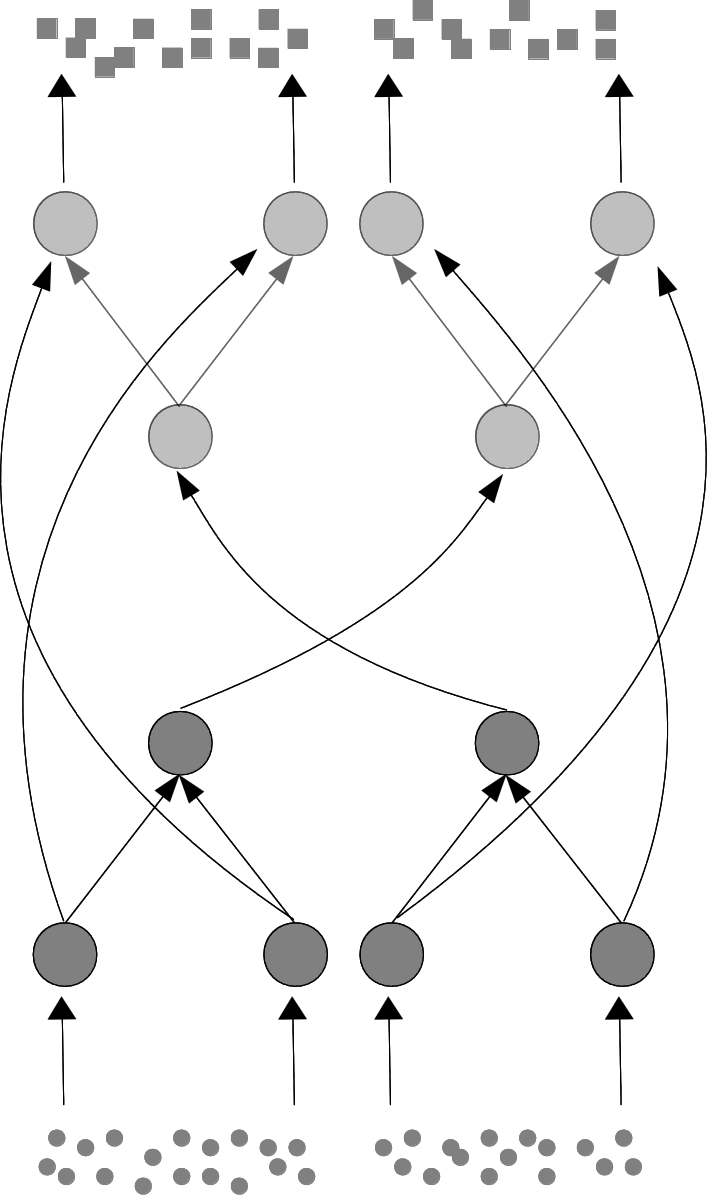}
    \put(9, 131){\tiny L2P}
    \put(27,131){\tiny L2P}
    \put(51,131){\tiny L2P}
    \put(69,131){\tiny L2P}

    \put(15,112){\tiny L2L}
    \put(40,110){\tiny L2L}
    \put(72,105){\tiny L2L}

    \put(7,75){\tiny M2L}
    \put(37,77){\tiny M2L}
    \put(72,70){\tiny M2L}

    \put(13,35){\tiny M2M}
    \put(40,44){\tiny M2M}
    \put(63,35){\tiny M2M}

    \put(9, 14){\tiny P2M}
    \put(26,14){\tiny P2M}
    \put(51,14){\tiny P2M}
    \put(68,14){\tiny P2M}
    \end{overpic}
  }
  \hfill
  \subfloat[FMM task flow]{
    \label{fig:dag}
    \raisebox{2em}{
      \begin{overpic}[scale=.505]{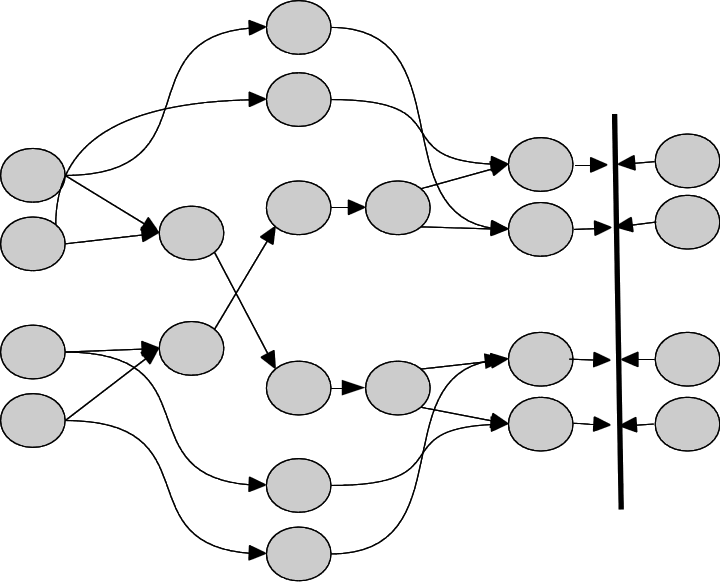}
        \put(02,30){\tiny P2M}
        \put(02,44){\tiny P2M}
        \put(02,65){\tiny P2M}
        \put(02,79){\tiny P2M}
        
        \put(33,68.5){\tiny M2M}
        \put(33,44.5){\tiny M2M}
        
        \put(55,3){\tiny M2L}
        \put(55,17){\tiny M2L}
        \put(55,36.5){\tiny M2L}
        \put(55,73){\tiny M2L}
        \put(55,95){\tiny M2L}
        \put(55,109){\tiny M2L}
        
        \put(76,37){\tiny L2L}
        \put(76,73){\tiny L2L}
        
        \put(104,30){\tiny L2P}
        \put(104,43){\tiny L2P}
        \put(104,69){\tiny L2P}
        \put(104,82){\tiny L2P}
        
        \put(133,30){\tiny P2P}
        \put(133,43){\tiny P2P}
        \put(133,70){\tiny P2P}
        \put(133,82){\tiny P2P}
      \end{overpic}
    }}
  \caption{Simplified FMM DAG.}
  \label{fig:fmmdag}
\end{figure}
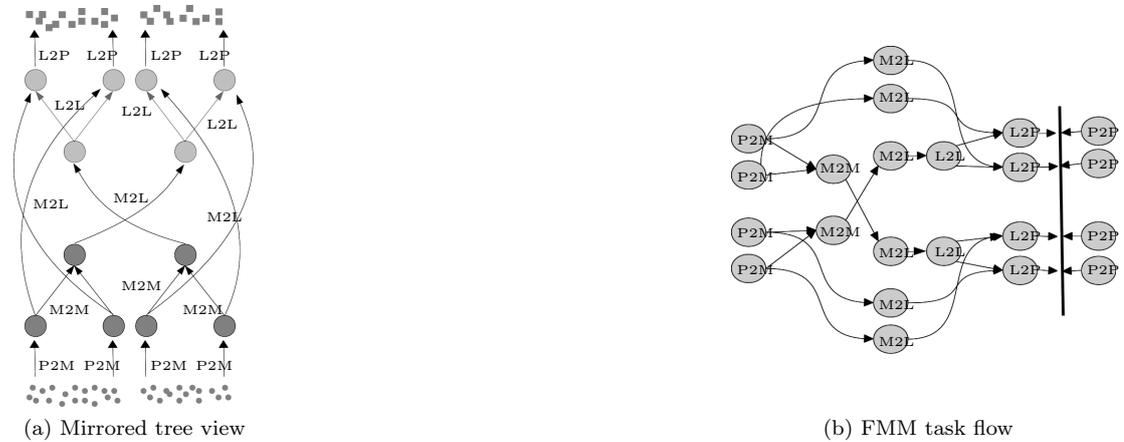

We start by subdividing the computational domain hierarchically into
subdomains, hereafter, referred to as cells. The resulting data structure is a
non-directed tree as presented in Figure~\ref{fig:unmirrored2}.
We let vertices represent cells (which contain data such as multipole or local
expansions) and edges represent the operations (P2P, P2M, L2L, M2M, L2P)
applied on them (see Figure~\ref{fig:unmirrored1}). The tree is traversed
twice. A bottom up traversal performs the P2M and M2M operations and a top
down traversal performs the M2L followed by the L2L and L2P operations. P2P
are usually associated to leaf cells and can be executed at any time except
when the corresponding L2P is being executed. M2L can be represented with
additional edges between siblings (see again
Figure~\ref{fig:unmirrored1}). However, the resulting graph is neither a tree
nor a DAG anymore since. The most straightforward way to derive a DAG from the
tree would consist of mirroring the tree as illustrated in
Figure~\ref{fig:mirrored}. The resulting DAG, where tasks are carried by
edges, can finally be turned into an appropriate DAG shown in
Figure~\ref{fig:dag} where tasks are carried by vertices (as required by the
runtime system, see Section~\ref{sec:starpu}).

\subsection{Pipelining the FMM DAG}

In a naive implementation of the FMM DAG each task works on one cell. For
example a P2P tasks computes the at most $26$ near-field interactions of one
cell at the leaf level, a M2M task computes the equivalent source values based
on the source values of its $8$ child cells or an M2L tasks computes the at
most $189$ far-field interactions. This approach leads to
\begin{itemize}
\item a large number of tasks (there are at most $8^\nu$ cells at level $\nu$,
  hence, the same number of tasks for all kernels)
\item a large number of dependencies (a cell has at most $189$ far-field
  interactions or a leaf cell has at most $26$ near-field interactions and all
  of them must be made available to each task)
\item a very small granularity (depends on $Acc$)
\end{itemize}

\section{Homogeneous case}
\label{sec:homogeneous}

In the previous section, we have shown how the FMM can be turned into a task
flow that can be executed by a runtime system.  Here, we consider multicore
platforms (Section~\ref{sec:homogeneous-setup}) and design a variant of the
task flow for which we tune the granularity of the tasks in order to
efficiently exploit homogeneous multicore architectures. We show that very
high parallel efficiency is achieved, which we explain thanks to a theoretical
breakdown of the computational costs.

\subsection{Experimental setup}
\label{sec:homogeneous-setup}

We consider two homogeneous platforms for assessing our algorithms.  The first
platform is composed of four deca-core Intel Xeon E7-4870 processors running
at $\unit[2.40]{GHz}$ (40 cores total). This machine is cache-coherent with
Non Uniform Memory Access (ccNUMA), which means that each core can access the
memory available on all sockets, with a higher latency if the accessed data is
on another socket. Each socket has 256~MB of random access memory (RAM) and a
30~MB L3 cache. Each CPU core has its own L1 and L2 caches of size 32~KB and
256~KB, respectively.

The second machine is an SGI Altix UV 100. It is also a ccNUMA machine,
composed of twenty octa-core Intel Xeon E7-8837 processors running at
$\unit[2.67]{GHz}$ (160 cores total). Each socket has 32~GB of RAM and 24~MB
of L3 cache. Each CPU core has its own L1 and L2 caches of size 32~KB and
256~KB, respectively.

In the rest of the paper, we will refer to these platforms as the \emph{four
  deca-core Intel Xeon E7-4870} and \emph{twenty octa-core Intel Xeon E7-8837}
machines, respectively.

\subsection{Pipelining strategies}
\label{sec:homogeneous-pipeline}

\begin{figure}[htbp]
\centering
\includegraphics[width=.45\textwidth]{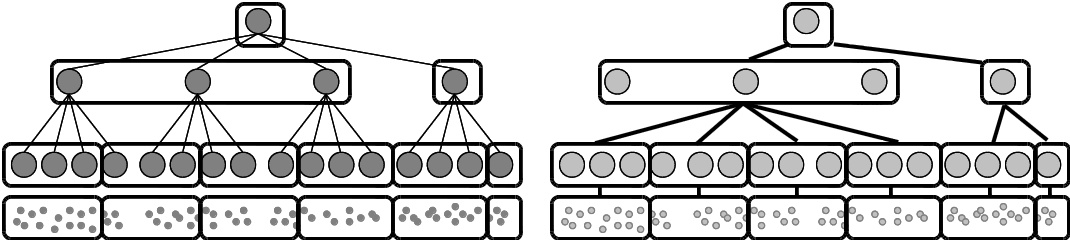}
\caption{A tree split in groups of three elements}
\label{fig_groups}
\end{figure}

The task flow proposed in Section~\ref{sec:fmmruntime} works on extremely fine
grained data which induces a large amount of tasks. The overhead for the
runtime system to handle all these tasks may become non negligible and produce
a large penalty on the total execution time. Preliminary experiments (not
reported here) showed that this approach does not scale. Assume, $t_i$ is the
time needed to schedule a task, $t_e$ the time to execute a task and $n_p$ the
number of available threads. If $t_e < n_pt_i$ then not sufficient work for
all threads will be available. We overcome this problem by increasing the
granularity of the tasks by letting them operate on groups of cells at the
same level as shown in Figure~\ref{fig_groups}.
On one hand, an increased task granularity leads to a higher performance of
the FMM kernels. On the other hand, fewer tasks are available and that limits
the concurrency.  Thus, we parameterize the granularity which allows us to
find a trade-off between performance and concurrency. The parameter is the
size $n_g$ of the group of cells we let a task work on. With this strategy we
obtain relatively regular tasks even for very irregular problems. Moreover, it
reduces the overall number of dependencies (see Fig.~\ref{fig:speedgroupsize}
for some studies). We use Morton ordering~\cite{Warren:1993} to group
cells. That is why they tend to be grouped also locally and, hence, they very
likely share common near and far-field interactions.

\subsection{Breakdown of the computational work}
\label{sec:quantitative_breakdown}
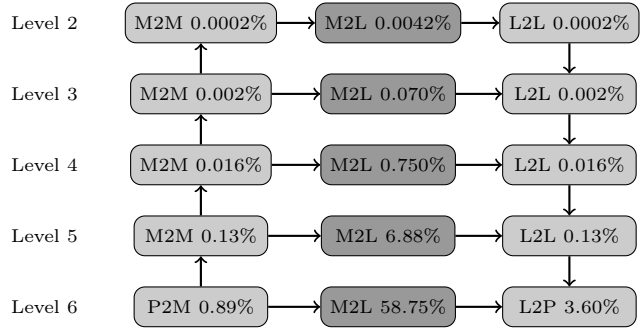
\begin{figure}[htbp]
  \centering
  \begin{tikzpicture}
    [font=\scriptsize,
    block/.style = {rectangle, draw=black, fill=black!20,
      align=center, rounded corners,
      minimum width=5em,
      minimum height=1.5em},
    blm2l/.style = {rectangle, draw=black, fill=black!40,
      align=center, rounded corners,
      minimum width=5em,
      minimum height=1.5em},
    line/.style = {draw, thick, ->}]
    \matrix [column sep=5mm,row sep=4mm]
    {
      \node (lev_2) {Level $2$}; &
      \node[block] (M2M_2) {M2M $0.0002\%$}; & 
      \node[blm2l] (M2L_2) {M2L $0.0042\%$}; &
      \node[block] (L2L_2) {L2L $0.0002\%$}; \\
      \node (lev_3) {Level $3$}; &
      \node[block] (M2M_3) {M2M $0.002\%$}; & 
      \node[blm2l] (M2L_3) {M2L $0.070\%$}; &
      \node[block] (L2L_3) {L2L $0.002\%$}; \\
      \node (lev_4) {Level $4$}; &
      \node[block] (M2M_4) {M2M $0.016\%$}; & 
      \node[blm2l] (M2L_4) {M2L $0.750\%$}; &
      \node[block] (L2L_4) {L2L $0.016\%$}; \\
      \node (lev_5) {Level $5$}; &
      \node[block] (M2M_5) {M2M $0.13\%$}; & 
      \node[blm2l] (M2L_5) {M2L $6.88\%$}; &
      \node[block] (L2L_5) {L2L $0.13\%$}; \\
      \node (lev_6) {Level $6$}; &
      \node[block] (P2M_6) {P2M $0.89\%$}; & 
      \node[blm2l] (M2L_6) {M2L $58.75\%$}; &
      \node[block] (L2P_6) {L2P $3.60\%$}; \\
    };
    \begin{scope}[every path/.style=line]
      \path (P2M_6) -- (M2M_5);
      \path (M2M_5) -- (M2M_4);
      \path (M2M_4) -- (M2M_3);
      \path (M2M_3) -- (M2M_2);
      \path (M2M_2) -- (M2L_2);
      \path (M2M_3) -- (M2L_3);
      \path (M2M_4) -- (M2L_4);
      \path (M2M_5) -- (M2L_5);
      \path (P2M_6) -- (M2L_6);
      \path (M2L_2) -- (L2L_2);
      \path (M2L_3) -- (L2L_3);
      \path (M2L_4) -- (L2L_4);
      \path (M2L_5) -- (L2L_5);
      \path (M2L_6) -- (L2P_6);
      \path (L2L_2) -- (L2L_3);
      \path (L2L_3) -- (L2L_4);
      \path (L2L_4) -- (L2L_5);
      \path (L2L_5) -- (L2P_6);
    \end{scope}
  \end{tikzpicture}
  \caption{The breakdown of the computational work of the FMM kernels
    ($N=20\cdot10^6$ uniformly in the unit-cube distributed particles, $h=7$,
    $Acc=7$) shows dependencies of the kernels at different levels and the
    percentage of the overall work which sums up to~$2.71\cdot10^{12}$
    floating point operations (P2P $28.75\%$)}
  \label{fig:fmm_macro_dag}
\end{figure}

In Figure \ref{fig:fmm_macro_dag} we present the breakdown of the
computational work for the different FMM kernels based on an example with
$N=20\cdot10^6$ uniformly distributed particles, an octree of height $h=7$ and
an accuracy $Acc=7$. It shows the level-wise grouped far-field operators (P2M,
M2M, M2L, L2L and L2P) and the respective number of required floating point
operations. Edges indicate dependencies: The algorithm starts with the P2M at
level $6$ (the leaf level), once it is finished it gives the go to the M2L at
the leaf level and the M2M at level $5$, and so on. Evidently, the work of the
M2L kernel is predominant.  Moreover, most of the work is done by the kernels
at the leaf level. Indeed, it decreases exponentially (we have a uniform
octree; hence, by a factor of $8$ per level).

\subsection{Parallel efficiency}	
\label{sec:homogeneous-scalability}

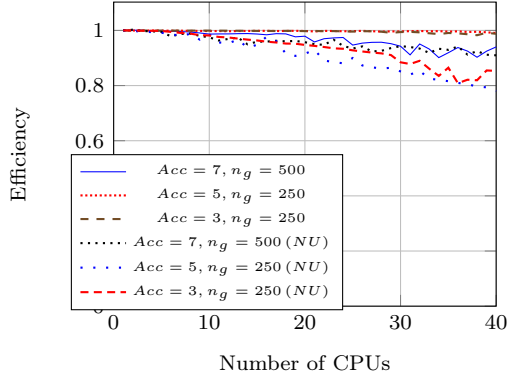
\begin{figure}[htbp]
  \centering
  \begin{tikzpicture}[font=\footnotesize]
    \begin{axis}[
      no markers,
      height=.27\textheight,
      width=.45\textwidth,
      xmin=0,
      xmax=40,
      ymin=0,
      xlabel=Number of CPUs,
      ylabel=Efficiency,
      legend style={at={(.6,.5)}, font=\tiny},
      grid=major,
      ytick={0,0.2,.4,.6,.8,1},
      ]
      \addplot+[solid]
      table[x index=0, y index=2] {Results/cpu_eff_riri_40.txt};
      \addlegendentry{$Acc=7,n_g=500$};
      \addplot+[densely dotted,thick]
      table[x index=0, y index=6] {Results/cpu_eff_riri_40.txt};
      \addlegendentry{$Acc=5,n_g=250$};
      \addplot+[dashed,thick]
      table[x index=0, y index=10] {Results/cpu_eff_riri_40.txt};
      \addlegendentry{$Acc=3,n_g=250$};
      \addplot+[dotted,thick]
      table[x index=0, y index=4] {Results/cpu_eff_riri_40.txt};
      \addlegendentry{$Acc=7,n_g=500\,(NU)$};
      \addplot+[loosely dotted,thick]
      table[x index=0, y index=8] {Results/cpu_eff_riri_40.txt};
      \addlegendentry{$Acc=5,n_g=250\,(NU)$};
      \addplot+[densely dashed,thick]
      table[x index=0, y index=12] {Results/cpu_eff_riri_40.txt};
      \addlegendentry{$Acc=3,n_g=250\,(NU)$};
    \end{axis}
  \end{tikzpicture}
  \caption{Parallel efficiency on up to $40$ CPUs on the \emph{four deca-core
      Intel Xeon E7-4870} machine for $20\cdot10^6$ particles (uniform
    distribution with $h=7$ and non-uniform (NU) with $h=8$); the granularity
    of tasks is parameterized by $n_g$, the number of cells it works on.}
  \label{fig:eff_cpu}
\end{figure}

\begin{figure}[htbp]
  \centering
  \begin{tikzpicture}[font=\footnotesize]
    \begin{axis}[
      no markers,
      height=.27\textheight,
      width=.45\textwidth,
      xmin=0,
      xmax=160,
      ymin=0,
      xlabel=Number of CPUs,
      ylabel=Efficiency,
      legend style={at={(.52,.55)}, font=\tiny},
      grid=major,
      ytick={0,0.2,.4,.6,.8,1}
      ]
      \addplot+[solid]
      table[x index=0, y index=2] {Results/starpu_efficiency_160.txt};
      \addlegendentry{$N=200\cdot10^6,Acc=7$};
      \addplot+[densely dotted,thick]
      table[x index=0, y index=6] {Results/starpu_efficiency_160.txt};
      \addlegendentry{$N=200\cdot10^6,Acc=5$};
      \addplot+[dashed,thick]
      table[x index=0, y index=10] {Results/starpu_efficiency_160.txt};
      \addlegendentry{$N=200\cdot10^6,Acc=3$};
      \addplot+[dotted,thick]
      table[x index=0, y index=4] {Results/starpu_efficiency_160.txt};
      \addlegendentry{$N=20\cdot10^6,Acc=7$};
      \addplot+[loosely dotted,thick]
      table[x index=0, y index=8] {Results/starpu_efficiency_160.txt};
      \addlegendentry{$N=20\cdot10^6,Acc=5$};
      \addplot+[densely dashed,thick]
      table[x index=0, y index=12] {Results/starpu_efficiency_160.txt};
      \addlegendentry{$N=20\cdot10^6,Acc=3$};
    \end{axis}
  \end{tikzpicture}
  \caption{Parallel efficiency on up to $160$ CPUs on the \emph{twenty
      octa-core Intel Xeon E7-8837} machine for uniformly distributed
    particles (for $N=20\cdot10^6$ we use $n_g=500$ and $h=7$, for
    $N=200\cdot10^6$ we use $n_g=1000$ and $h=8$); the granularity of tasks is
    parameterized by $n_g$, the number of cells it works on.}
  \label{fig:eff_cpu_160}
\end{figure}
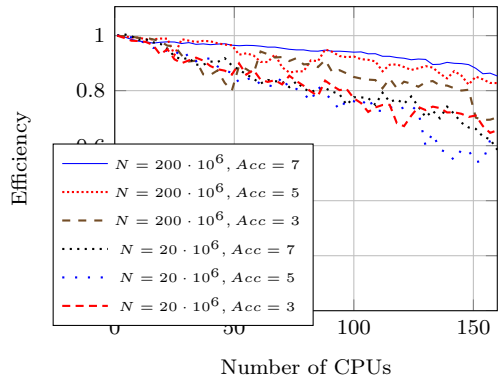

We now study the \emph{parallel efficiency} of our approach, defined as
\begin{equation*}
e_n=\frac{t_1}{nt_n},
\end{equation*} 
where $t_n$ denotes the measured execution time with $n$ computational devices.
In the Figures~\ref{fig:eff_cpu} and \ref{fig:eff_cpu_160} we present studies
obtained on the \emph{four deca-core Intel Xeon E7-4870} and \emph{twenty
  octa-core Intel Xeon E7-8837} machines, respectively. We present studies for
uniformly and non-uniformly distributed particles and for different accuracies
(defined as $(\ell,\varepsilon_{\text{SVD}})=(Acc,10^{-Acc})$). All studies
feature an extraordinary good scaling: between $80\%$ and $98\%$ efficiency at
$40$ CPUs in Figure~\ref{fig:eff_cpu} and between $60\%$ and $86\%$ efficiency
at $160$ CPUs in Figure~\ref{fig:eff_cpu_160}.
In order to understand the scaling behaviour more in detail we need to analyze
the influence of the granularity of the tasks to be scheduled by the runtime
system. In Section~\ref{sec:homogeneous-pipeline} we have parameterized the
granularity of tasks by means of the number of cells $n_g$ they work on (see
Fig.~\ref{fig:speedgroupsize} for some studies). Obviously, efficiency is
related to concurrency, which can be increased by adding another level to the
tree or by reducing the granularity of the tasks. In the uniform case in
Figure~\ref{fig:eff_cpu} the high accuracy example is less efficient than the
other two. The reason is that it does not provide enough concurrency due to a
larger granularity ($n_g=500$ compared to $250$). In the non-uniform case the
high accuracy example is the most efficient one. We have added one more level
to the tree, hence, there is enough concurrency even for tasks of larger
granularity.
In Figure~\ref{fig:eff_cpu_160} we kept the granularity constant. The
concurrency does not change but the granularity changes depending on the
chosen accuracy (higher accuracy leads to larger granularity). All studies
behave as expected, those with larger granularity are more efficient.

\begin{figure}[htbp]
\centering
\includegraphics[width=.48\textwidth,height=.35\textheight]
{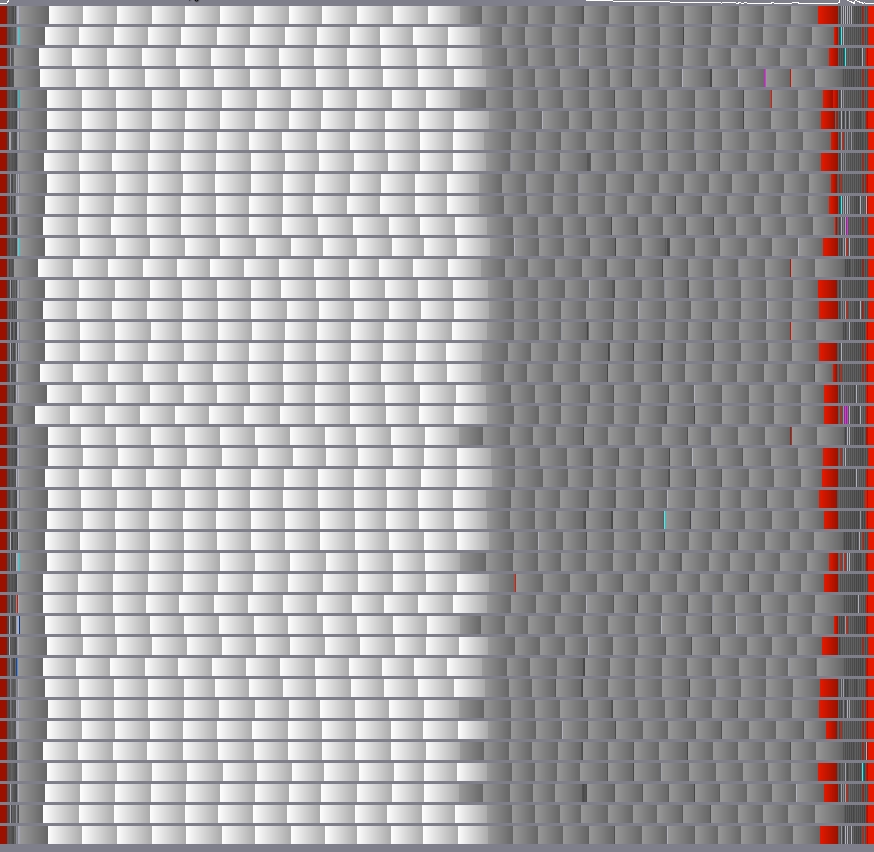}
\caption{Trace for the study in Figure~\ref{fig:eff_cpu} with $N=20\cdot10^6$
  and $Acc=7$ (execution time $t=\unit[23]{s}$).}
\label{fig:no_priorities}
\end{figure}

Figure~\ref{fig:no_priorities} shows the execution trace obtained by the
runtime system StarPU. Each horizontal lane shows the occupancy of a
particular processing unit (here $40$ CPUs) as a function of the time.
\begin{figure}[htbp]
  \centering
  \raisebox{1.3em}{
    \begin{overpic}[scale=.5]{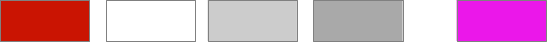}
      \put(10,-13){\footnotesize idle}
      \put(51,-13){\footnotesize P2P}
      \put(88,-13){\footnotesize M2L}
      \put(115,-13){\footnotesize other kernels}
      \put(168,-13){\footnotesize copying data}
    \end{overpic}}
  \caption{Color legend for traces}
  \label{fig:legend}
\end{figure}
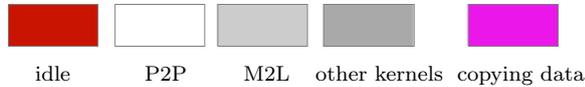
As illustrated in Figure~\ref{fig:legend}, red (dark) sections denote idle
time, white sections denote the P2P kernel, light gray sections denote the M2L
kernel and medium gray sections denote all other kernels (P2M, M2M, L2L, L2P;
as can be seen in Figure~\ref{fig:fmm_macro_dag} their work share is
vanishingly small compared to P2P and M2L). The purple color denotes that a
tasks is available but not the data. The trace shows a highly pipelined
execution. Depending on the execution, a barrier can be observed before the
L2P. This is due to data dependencies in the L2L meanly at leaves level.




\section{Heterogeneous case}
\label{sec:heterogeneous}

\begin{figure}[htbp]
  \centering
  \begin{tikzpicture}[font=\footnotesize]
    \begin{loglogaxis}[
      height=.27\textheight,
      width=.48\textwidth,
      xlabel=Number of interactions,
      ylabel=Time $(\unit{s})$,
      legend style={at={(.62,1.1)}, font=\tiny},
      legend columns=2,
      grid=both,
      minor x tick num=5,
      ]
      \addplot[mark=square, blue, only marks]
      table[x index=0, y index=1] {Images/Calib/starpu_P2P_avg.data};
      \addlegendentry{Measured CPU Time};
      \addplot[solid, blue, thick, domain=1e2:1e4, samples=40]
      table[x index=0, y index=1] {Images/Calib/calib.txt};
      \addlegendentry{Estimated CPU Time};
      \addplot[mark=pentagon, red, only marks]
      table[x index=0, y index=3] {Images/Calib/starpu_P2P_avg.data};
      \addlegendentry{Measured GPU-1 Time};
      \addplot[densely dotted, thick, red, domain=1e2:1e4, samples=40]
      table[x index=0, y index=2] {Images/Calib/calib.txt};
      \addlegendentry{Estimated GPU-1 Time};
      \addplot[mark=triangle, green, only marks]
      table[x index=0, y index=5] {Images/Calib/starpu_P2P_avg.data};
      \addlegendentry{Measured GPU-2 Time};
      \addplot[dashed, thick, green, domain=1e2:1e4, samples=40]
      table[x index=0, y index=3] {Images/Calib/calib.txt};
      \addlegendentry{Estimated GPU-2 Time};
      \addplot[mark=diamond, only marks]
      table[x index=0, y index=7] {Images/Calib/starpu_P2P_avg.data};
      \addlegendentry{Measured GPU-3 Time};
      \addplot[dotted, thick, domain=1e2:1e4, samples=40]
      table[x index=0, y index=4] {Images/Calib/calib.txt};
      \addlegendentry{Estimated GPU-3 Time};
    \end{loglogaxis}
  \end{tikzpicture}
  \caption{Calibration results of the \emph{Heft} scheduler after two
    simulations for the CPU and the three GPU.}
  \label{fig:P2PCalibration}
\end{figure}
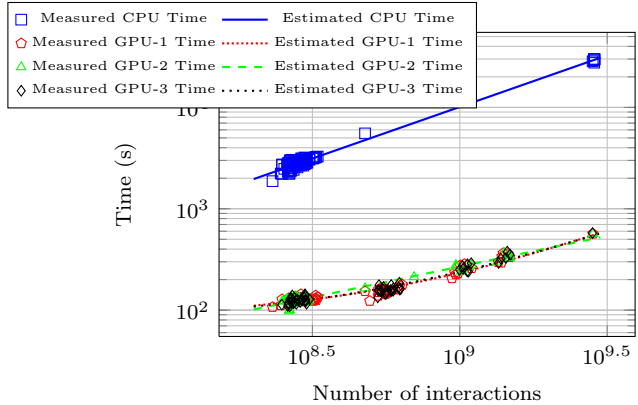

If GPUs are now available on the computational node, the execution of a task
can be deported on them by the runtime system (see
Section~\ref{sec:starpu}). As discussed above, most FMM codes for
heterogeneous machines decide how to distribute the work load before starting
the actual parallel execution. With such a static approach, it is very hard to
consistently achieve efficient load balancing. StarPU allows for a dynamic
distribution of the work load based on given scheduling strategies
(see~\cite{AugThiNamWac11CCPE}). We use the \emph{Eager} scheduler for all
computations. Other schedulers are available on StarPU, but they are not yet
considered as stable. Some tests have been performed with the \emph{Heft}
scheduler which is able to estimate and use the duration of tasks to schedule
them optimally. Such estimation is based on preliminary runs where the
duration of tasks is stored in a database. In Fig.~\ref{fig:P2PCalibration}
calibration results for the \emph{Heft} scheduler of the P2P kernel are
presented. During two preliminary runs the duration of all P2P tasks as a
function of the number of interactions are stored in a database and based
thereon the scheduler is able to estimate the duration of tasks in future
runs.

In Section~\ref{sec:heterogeneous-intermediatenearfield}, we consider
executions where near-field (P2P) and far-field (other kernels) interactions
are well balanced. Relying on a high performance blocked P2P kernel, we show
that processing P2P (and only P2P) on GPUs allows for achieving well balanced
executions. We then consider the case where the near-field is dominant in
Section~\ref{sec:heterogeneous-nearfield} and we show that we continue to
achieve a good load balance by scheduling P2P tasks on demand on CPU or
GPU. Finally, we study the case where the far-field is dominant in
Section~\ref{sec:heterogeneous-farfield}. By relying on a M2L GPU kernel (not
yet optimized), we manage to maintain a high pipeline and well balanced
execution. 

%

\subsection{Experimental setup}
\label{sec:heterogeneous-setup}

We consider a dual-socket hexa-core host machine based on Intel X5650 Nehalem
processors operating at $2.67$\,GHz. Each socket has $12$\,MB of L3 cache and
each core has $256$\,KB of L2 cache. The size of the main memory is
$48$\,GB. Moreover, we have two different GPU sets associated to it. They
consist of three Nvidia M$2070$, respectively, M$2090$ Fermi accelerators and
are connected to the host with a 16x PCI bus. Each GPU has $6$\,GB of GDDR-5
of memory and $14$ processors ($448$, respectively $512$ cores), operating at
$1.15$\,GHz. In the rest of the paper, we will refer to this platforms as the
\emph{Nehalem-Fermi (M2070)}, respectively, \emph{Nehalem-Fermi (M2090)}
machine. If we do not need to explicitely refer to one of these clusters we
simply call them heterogeneous machines since they are composed of CPUs and
GPUs. Because StarPU dedicates a CPU core in order to handle a GPU (see Section~\ref{sec:starpu}), both machine will be viewed as a nine CPU cores node enhanced with three GPUs.

\subsection{Balancing near and far-field}
\label{sec:heterogeneous-intermediatenearfield}

We consider executions where near-field (P2P) and far-field (other kernels)
computational costs are well balanced. Relying on a high performance blocked
P2P kernel, we design a first scheme that processes all P2P (and only P2P) on
GPUs and performs highly pipelined executions.

We have derived from~\cite{takahashi:2012} a GPU kernel that operates on
groups of $n_g$ cells. Figure~\ref{fig:P2PCalibration} compares the execution
times of our CPU and GPU P2P kernels. Because we group multiple cells to be
processed into a single kernel call, not all particles interact directly with
each other. The time spent for executing this kernel is thus neither
proportional to the number of cells nor
particles. Figure~\ref{fig:P2PCalibration} indeed shows that the execution
time of the kernel is proportional to the number of effective interactions
(x-axis).

Our main goal is to obtain perfectly piplined executions. Hence, we choose the
reference example settings such that the computational load is well balanced
between the three available GPUs and nine available CPUs. Initially we process
the near-field (P2P kernels) on the GPUs and the far-field (all other kernels)
on the CPUs. In order to end up with a roughly matching execution time we
choose $N=38\cdot10^6$, $Acc=5$ and $h=7$. Moreover, we chose a large block
size ($n_g=1500$) in order to achieve a high flop-rate for the P2P GPU kernel
($\unitfrac[102]{GFlop}{s}$).

\begin{figure}[htbp]
\centering
\includegraphics[width=.48\textwidth,height=.23\textheight]
{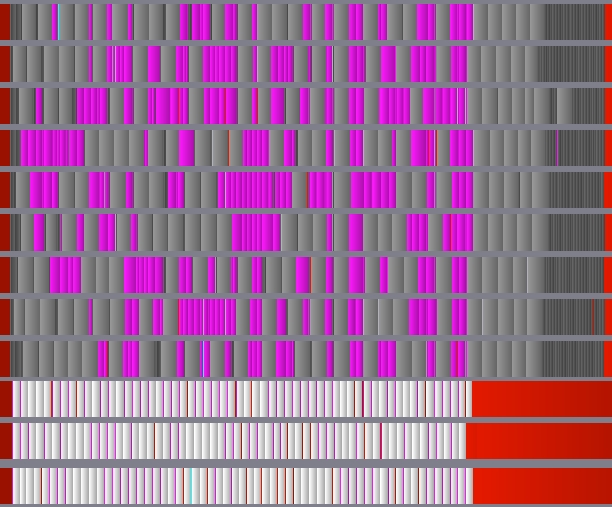}
\caption{Execution trace on the \emph{Nehalem-Fermi (M2070)} heterogeneous
  machine ($N=38\cdot10^6$, $Acc=5$, $h=7$, $n_g=1500$) with balanced near and
  far-field. Each P2P task is followed by the P2Preduce tasks. The execution
  time is $t=\unit[18.50]{s}$. The nine first lanes represent CPU cores
  occupancy and the three last ones GPUs.}
\label{fig:GpuNotGood}
\end{figure}

Figure~\ref{fig:GpuNotGood} shows the execution trace of the task flow
proposed in Section~\ref{sec:homogeneous} on \emph{Nehalem-Fermi (M2070)}, the
heterogeneous machine presented in Section~\ref{sec:heterogeneous-setup}. We
recall that for the moment we force all P2P tasks to be executed on GPUs. As
expected, they are perfectly pipelined (last three lanes in
Figure~\ref{fig:GpuNotGood}) since they are all independent one from
another. However, CPU cores (first nine lanes) are not fully occupied (purple
color means that a task is available but its data not). The reason is that P2P
tasks, executed on GPUs, are followed by P2Preduce tasks which are executed on
CPUs. What is the P2Preduce tasks for? It basically exploits the fact that if
source and target particles are the same the resulting matrix becomes
symmetric. Hence, we do not only write the resulting potential and forces on
the target particles but also on source particles. The P2Preduce tasks perform
this data reduction and cannot be executed at the same time as the
corresponding L2P tasks.

\begin{figure*}[htbp]
  \centering
  \subfloat[\emph{Nehalem-Fermi (M2070)} $(t=14.78\,\text{s})$]{
    \label{fig:fermi}
    \includegraphics[width=.48\textwidth,height=.23\textheight]
    {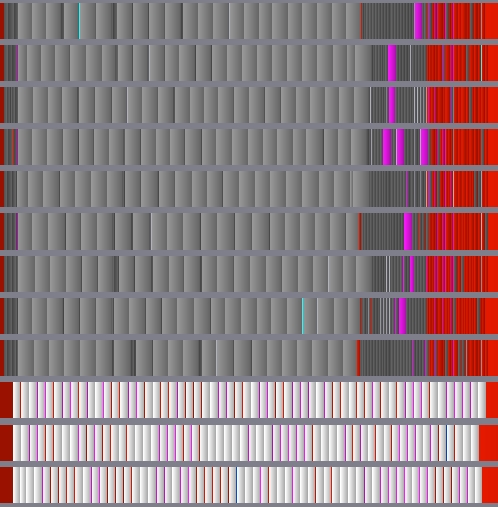}
  }
  \hfill
  \subfloat[\emph{Nehalem-Fermi (M2090)} $(t=13.34\,\text{s})$]{
    \label{fig:fermi2090}
    \includegraphics[width=.48\textwidth,height=.23\textheight]
    {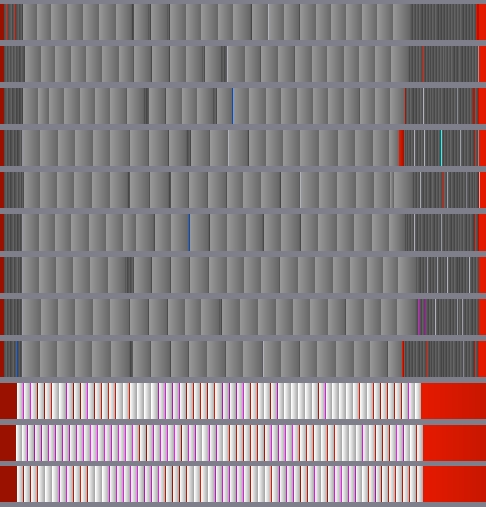}
  }
  \caption{Execution trace on both heterogeneous machines ($N=38\cdot10^6$,
    $Acc=5$, $h=7$, $n_g=1500$) with balanced near and far-field. Data from
    GPU are prefetched and P2Preduce tasks are performed last. The nine first
    lanes represent CPU cores occupancy and the three last ones GPUs.}
  \label{fig:GpuGood}
\end{figure*}

Finally, by forcing P2Preduce tasks to be processed after L2P tasks and
prefetching all data movements between CPUs and GPUs, we obtain a perfect
pipeline (see Figure~\ref{fig:GpuGood}). It reduces the walltime from
$\unit[18.50]{s}$ to $\unit[14.78]{s}$.

\subsection{Dominant near-field}
\label{sec:heterogeneous-nearfield}

\begin{figure}[htbp]
\centering
\includegraphics[width=.48\textwidth,height=.23\textheight]
{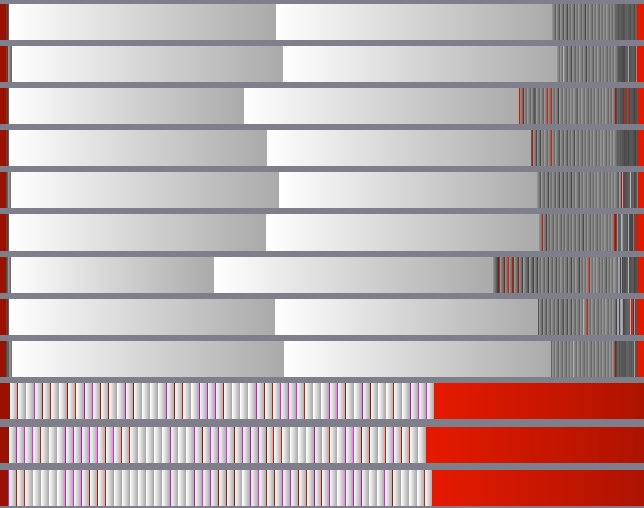}
\caption{Execution trace on the \emph{Nehalem-Fermi (M2070)} heterogeneous
  machine with a dominant near-field ($N=38\cdot10^6$, $Acc=3$, $h=7$,
  $n_g=1500$). The execution time is $t=\unit[19.12]{s}$. The nine first lanes
  represent CPU cores occupancy and the three last ones GPUs.}
\label{fig:BeforeP2PCalibration}
\end{figure}

Next, we consider an execution with a lower accuracy ($Acc=3$). M2L
computational cost is reduced (M2L is the most sensitive kernel to the
accuracy) and therefore P2P strongly dominates.  If P2P is fully processed on
GPU, CPUs remain idle most of the time (trace not reported here). We therefore
allow for scheduling dynamically P2P on CPU or GPU. In order ensure
prefetching, the scheduling decision has to be taken ahead of time (so that
data have time to be moved between the different computational
units). However, because of the heterogeneity of the computational power of
CPUs and GPUs, this may still lead to unbalanced executions
(Figure~\ref{fig:BeforeP2PCalibration}. By limiting the anticipation of such
decisions, we managed to balance correctly the load (trace not reported here).


\subsection{Dominant far-field}
\label{sec:heterogeneous-farfield}

\begin{figure*}[htbp]
  \centering
  \subfloat[No M2L tasks on GPU $(t=49.02\,\text{s})$]{
    \label{fig:M2LnoGPU}
    \includegraphics[width=.48\textwidth,height=.23\textheight]
    {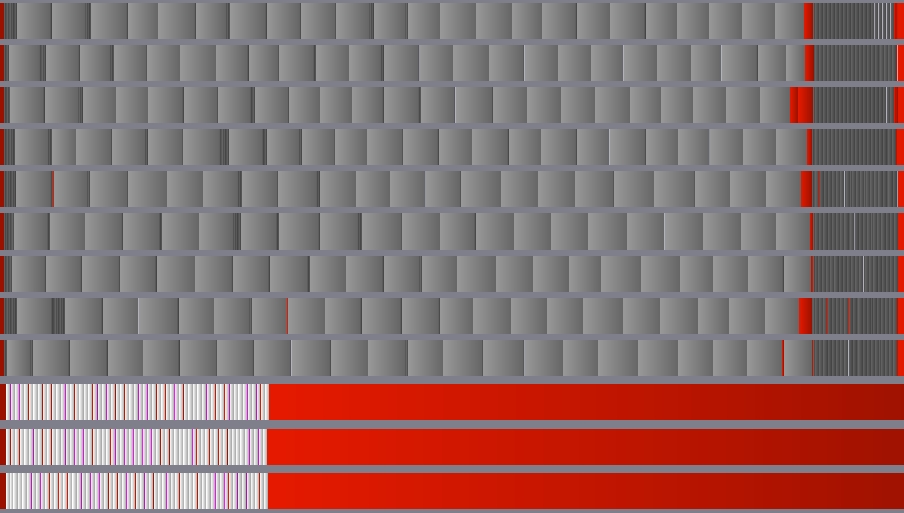}
  }
  \hfill
  \subfloat[Some M2L tasks on GPU $(t=42.09\,\text{s})$]{
    \label{fig:M2LwithGPU}
    \includegraphics[width=.48\textwidth,height=.23\textheight]
    {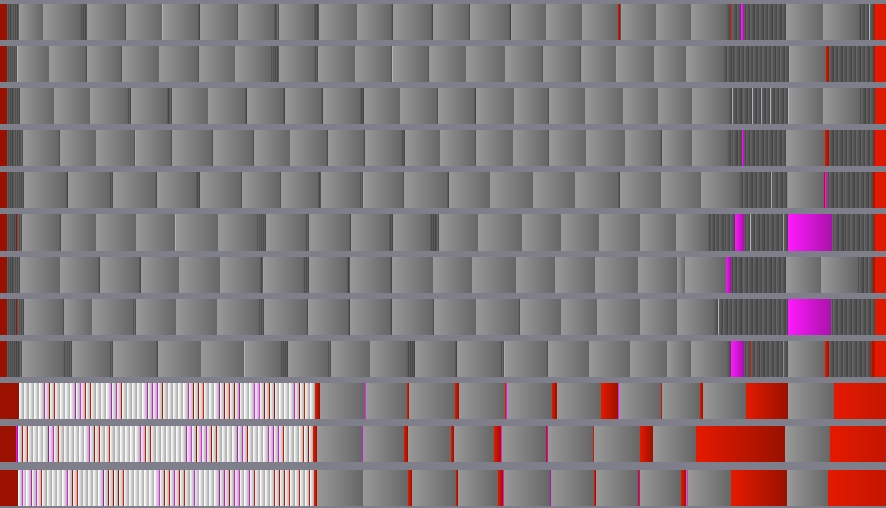}
  }
  \caption{Execution trace on \emph{Nehalem-Fermi (M2070)} with a dominant
    far-field ($N=38\cdot10^6$, $Acc=7$, $h=7$, $n_g=1500$). The nine first
    lanes represent CPU core occupancy and the three last ones the GPU core
    occupancy.}
  \label{fig:M2LGPU}
\end{figure*}

We finally, by increasing the accuracy ($Acc=7$), we consider the case where
the far-field dominates. Since M2L tasks become dominant, the computation is
unbalanced (Figure~\ref{fig:M2LnoGPU}). Thus, we have designed a (non
optimized) GPU M2L kernel in order to show that we can efficiently balance the
load by scheduling M2L tasks on demand (P2P tasks are processed on GPUs).
Figure~\ref{fig:M2LwithGPU} shows the resulting trace. The pipeline is again
very efficient and the load well balanced (but the performance of the M2L
kernel would remain to be optimized).

\section{Performance benchmarks}
\label{sec:finalresults}

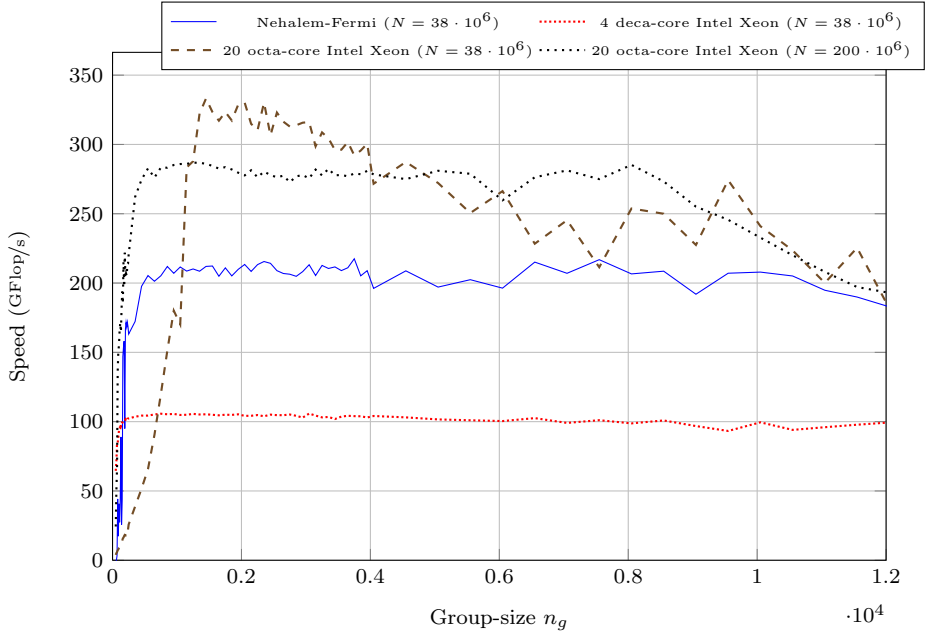
\begin{figure}[htbp]
  \centering
  \begin{tikzpicture}[font=\footnotesize]
    \begin{axis}[
      no markers,
      height=.4\textheight,
      width=.8\textwidth,
      xmin = 0,
      ymin = 0,
      xmax = 12000,
      xlabel=Group-size $n_g$,
      ylabel=Speed $(\unitfrac{GFlop}{s})$,
      legend columns=2,
      legend style={at={(1.05,1.1)}, font=\tiny},
      grid=major,
      ]
      \addplot+[solid]
      table[x index=0, y index=5] {Traces/BlockSize/result.txt};
      \addlegendentry{Nehalem-Fermi $(N=38\cdot10^6)$};
      \addplot+[densely dotted,thick]
      table[x index=0, y index=6] {Traces/BlockSize/result.txt};
      \addlegendentry{$4$ deca-core Intel Xeon $(N=38\cdot10^6)$};
      \addplot+[dashed,thick]
      table[x index=0, y index=7] {Traces/BlockSize/result.txt};
      \addlegendentry{$20$ octa-core Intel Xeon $(N=38\cdot10^6)$};
      \addplot+[dotted,thick]
      table[x index=0, y index=8] {Traces/BlockSize/result.txt};
      \addlegendentry{$20$ octa-core Intel Xeon $(N=200\cdot10^6)$};
    \end{axis}
  \end{tikzpicture}
  \caption{Flop-rates on three different computers for different group sizes
    $n_g<12000$ (particles per group) for $Acc=5$ and $h=7$ (uniform particle
    distribution)}
  \label{fig:speedgroupsize}
\end{figure}

Several parameters influence the achieved flop-rates. One is the group-size
$n_g$, but Fig.~\ref{fig:speedgroupsize} shows that its influence is less
critical. In other words, a broad range of $n_g$ leads to the same good
flop-rates. Interesting is the benchmark on the $20$ octa-core Intel Xeon
machine for $N=38\cdot10^6$: it is best performing for $1000<n_g<3000$. This
is a rather small interval compared to the other three benchmarks. The reason
is that larger $n_g$ does not provide enough parallelism anymore.

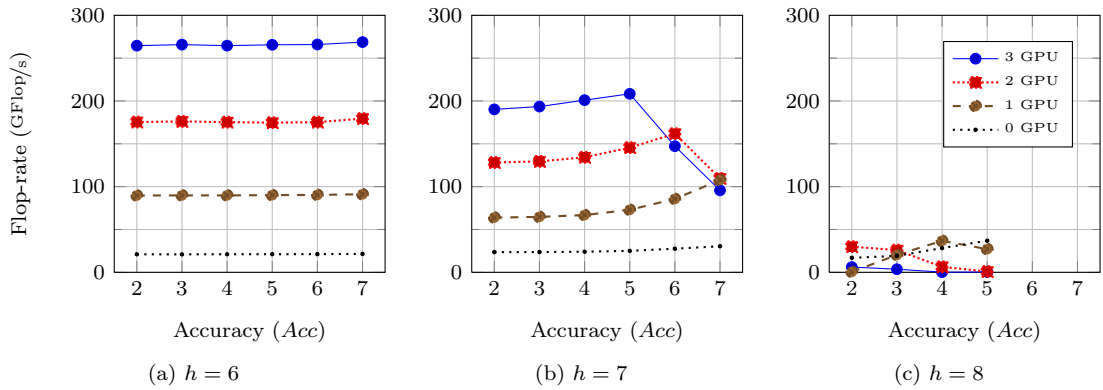
\begin{figure*}[htbp]
  \centering
  \subfloat[$h=6$]{
    \label{fig:unformh6}
    \begin{tikzpicture}[font=\footnotesize]
      \begin{axis}[
        height=.24\textheight,
        width=.35\textwidth,
        xlabel=Accuracy ($Acc$),
        xtick=data,
        ylabel=Flop-rate $(\unitfrac{GFlop}{s})$,
        minor y tick num=1,
        ymin=0,
        ymax=300,
        grid=both,
        ]
        \addplot+[solid]
        table[x index=0, y index=9] {finalplottxt/h6.38kk.txt};
        \addplot+[densely dotted,thick]
        table[x index=0, y index=8] {finalplottxt/h6.38kk.txt};
        \addplot+[dashed,thick]
        table[x index=0, y index=7] {finalplottxt/h6.38kk.txt};
        \addplot+[dotted,thick]
        table[x index=0, y index=6] {finalplottxt/h6.38kk.txt};
      \end{axis}
    \end{tikzpicture}
  }
  \hfill
  \subfloat[$h=7$]{
    \label{fig:uniformh7}
    \begin{tikzpicture}[font=\footnotesize]
      \begin{axis}[
        height=.24\textheight,
        width=.35\textwidth,
        xtick=data,
        xlabel=Accuracy ($Acc$),
        minor y tick num=1,
        ymin=0,
        ymax=300,
        grid=both,
        ]
        \addplot+[solid]
        table[x index=0, y index=9] {finalplottxt/h7.38kk.txt};
        \addplot+[densely dotted,thick]
        table[x index=0, y index=8] {finalplottxt/h7.38kk.txt};
        \addplot+[dashed,thick]
        table[x index=0, y index=7] {finalplottxt/h7.38kk.txt};
        \addplot+[dotted,thick]
        table[x index=0, y index=6] {finalplottxt/h7.38kk.txt};
      \end{axis}
    \end{tikzpicture}
  }
  \hfill
  \subfloat[$h=8$]{
    \label{fig:uniformh8}
    \begin{tikzpicture}[font=\footnotesize]
      \begin{axis}[
        height=.24\textheight,
        width=.35\textwidth,
        xlabel=Accuracy ($Acc$),
        legend style={at={(.9,.9)},font=\tiny},
        minor y tick num=1,
        grid=both,
        xtick={2,3,4,5,6,7},
        ymin=0,
        ymax=300,
        xmin=1.5,
        xmax=7.5,
        ]
        \addplot+[solid]
        table[x index=0, y index=9] {finalplottxt/h8.38kk.txt};
        \addlegendentry{$3$ GPU};
        \addplot+[densely dotted,thick]
        table[x index=0, y index=8] {finalplottxt/h8.38kk.txt};
        \addlegendentry{$2$ GPU};
        \addplot+[dashed,thick]
        table[x index=0, y index=7] {finalplottxt/h8.38kk.txt};
        \addlegendentry{$1$ GPU};
        \addplot+[dotted,thick]
        table[x index=0, y index=6] {finalplottxt/h8.38kk.txt};
        \addlegendentry{$0$ GPU};
      \end{axis}
    \end{tikzpicture}
  }
  \caption{Flop-rates for uniform distribution (in the unit-cube) of
    $38\cdot10^6$ particles and group size $n_g=1000$}
  \label{fig:uniformflops}
\end{figure*}

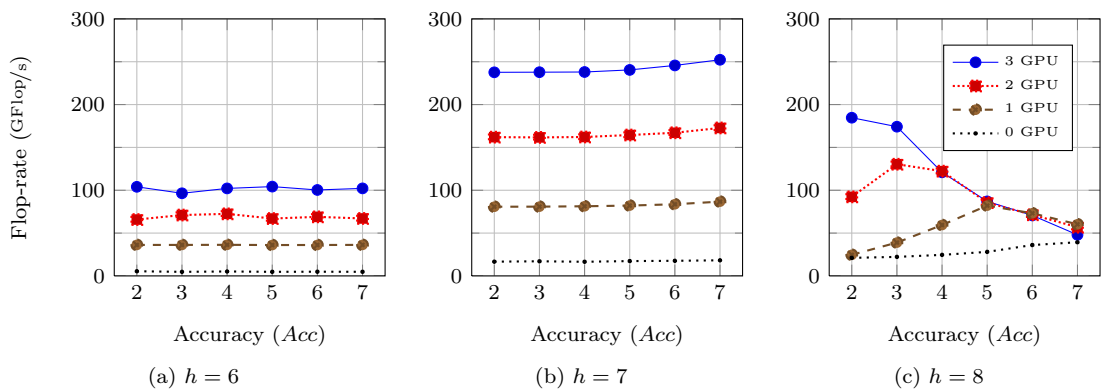
\begin{figure*}[htbp]
  \centering
  \subfloat[$h=6$]{
    \label{fig:nonunformh6}
    \begin{tikzpicture}[font=\footnotesize]
      \begin{axis}[
        height=.24\textheight,
        width=.35\textwidth,
        xlabel=Accuracy ($Acc$),
        ylabel=Flop-rate $(\unitfrac{GFlop}{s})$,
        ymin=0,
        xtick=data,
        ymax=300,
        minor y tick num=1,
        grid=both,
        ]
        \addplot+[solid]
        table[x index=0, y index=9] {finalplottxt/h6.s38kk.txt};
        \addplot+[densely dotted,thick]
        table[x index=0, y index=8] {finalplottxt/h6.s38kk.txt};
        \addplot+[dashed,thick]
        table[x index=0, y index=7] {finalplottxt/h6.s38kk.txt};
        \addplot+[dotted,thick]
        table[x index=0, y index=6] {finalplottxt/h6.s38kk.txt};
      \end{axis}
    \end{tikzpicture}
  }
  \hfill
  \subfloat[$h=7$]{
    \label{fig:nonuniformh7}
    \begin{tikzpicture}[font=\footnotesize]
      \begin{axis}[
        height=.24\textheight,
        width=.35\textwidth,
        xlabel=Accuracy ($Acc$),
        ymin=0,
        ymax=300,
        xtick=data,
        minor y tick num=1,
        grid=both,
        ]
        \addplot+[solid]
        table[x index=0, y index=9] {finalplottxt/h7.s38kk.txt};
        \addplot+[densely dotted,thick]
        table[x index=0, y index=8] {finalplottxt/h7.s38kk.txt};
        \addplot+[dashed,thick]
        table[x index=0, y index=7] {finalplottxt/h7.s38kk.txt};
        \addplot+[dotted,thick]
        table[x index=0, y index=6] {finalplottxt/h7.s38kk.txt};
      \end{axis}
    \end{tikzpicture}
  }
  \hfill
  \subfloat[$h=8$]{
    \label{fig:nonuniformh8}
    \begin{tikzpicture}[font=\footnotesize]
      \begin{axis}[
        height=.24\textheight,
        width=.35\textwidth,
        xlabel=Accuracy ($Acc$),
        legend style={at={(.9,.9)},font=\tiny},
        ymin=0,
        xtick=data,
        ymax=300,
        minor y tick num=1,
        grid=both,
        ]
        \addplot+[solid]
        table[x index=0, y index=9] {finalplottxt/h8.s38kk.txt};
        \addlegendentry{$3$ GPU};
        \addplot+[densely dotted,thick]
        table[x index=0, y index=8] {finalplottxt/h8.s38kk.txt};
        \addlegendentry{$2$ GPU};
        \addplot+[dashed,thick]
        table[x index=0, y index=7] {finalplottxt/h8.s38kk.txt};
        \addlegendentry{$1$ GPU};
        \addplot+[dotted,thick]
        table[x index=0, y index=6] {finalplottxt/h8.s38kk.txt};
        \addlegendentry{$0$ GPU};
      \end{axis}
    \end{tikzpicture}
  }
  \caption{Flop-rates for non-uniform distribution (on the unit-sphere) of
    $38\cdot10^6$ particles and group size $n_g=1000$}
  \label{fig:nonuniformflops}
\end{figure*}

In Fig.~\ref{fig:uniformflops} we show flop--rates studies for uniform (in the
unit-cube) and in Fig.~\ref{fig:nonuniformflops} for non-uniform (on the
unit-sphere) particle distributions. The number of particles is in either case
$N=38\cdot10^6$. We vary the octree height $h=6,7,8$. By doing so we can
redeploy work between near- and far-field. A smaller $h$ means more near- and
less far-field and a larger $h$ the other way around. Moreover, we vary the
accuracies $Acc=2,\dots,7$. By doing so, the cost for evaluating the
near-field does not change but higher $Acc$ increases the cost for evaluating
the far-field. All, these effects can obviously be recognized in the
figures~\ref{fig:uniformflops} and \ref{fig:nonuniformflops}.

In the figures~\ref{fig:unformh6}, \ref{fig:nonunformh6} and
\ref{fig:nonuniformh7} the computations are clearly dominated by the
near-field, i.e., the P2P kernels. Except in the $0$ GPU case they are all
performed on the GPU. In either case we observe the linear scaling for
additional GPUs. In the figures~\ref{fig:uniformh7} and \ref{fig:nonuniformh8}
we have a larger $h$ and after $Acc=5$, respectively, $Acc=3$ the computations
are dominated by the far-field, i.e., mainly the M2L kernels. They are
scheduled by StarPU dynamically to available CPUs and GPUs. However, since our
GPU implementation of the M2L kernel is not yet optimized, the flop-rate
breaks down once the far-field becomes dominant. In Fig.~\ref{fig:uniformh8}
we have the uniform case with $h=8$, which is clearly to large (about $8^7$
leafs which contain only about $18$ particles). The flop-rate is one order of
magnitude smaller then for $h=6,7$. In Fig.~\ref{fig:uniformh8} the examples
for $Acc=6,7$ have not been computed.

What is the main difference between uniform and non-uniform particle
distributions? In the uniform case particles are distributed in a
tree-dimensional subset of $\mathbb{R}^3$, in the non-uniform case they are
distributed in a two-dimensional subset of $\mathbb{R}^3$, only. This leads to
an octree in the uniform case and to a quadtree in the non-uniform case. Since
$N$ is the same and the convex-hull is of the same order in both cases, at a
given level of the tree a non-empty cluster in the non-uniform case tends to
contain an order of magnitude more particles compared to the uniform
case. That is why if we compare Fig.~\ref{fig:unformh6} and
Fig.~\ref{fig:nonuniformh7} we obtain the same flop-rates even though $h$ is
different.


\begin{figure*}[htbp]
  \centering
  \subfloat[$0$ GPU]{
    \label{fig:0gpu}
    \begin{tikzpicture}[font=\footnotesize]
      \begin{semilogyaxis}[
        height=.3\textheight,
        width=.5\textwidth,
        ylabel=Time $(\unit{s})$,
        xlabel=Accuracy $(Acc)$,
        minor y tick num=3,
        grid=both,
        xtick=data,
        ymin=0,
        ymax=1200,
        ymin=9,
        xmin=1.5,
        xmax=7.5,
        ]
        \addplot+[dashed,thick]
        table[x index=0, y index=1] {finalplottxt/h6.38kk.txt};
        \addplot+[dotted,thick]
        table[x index=0, y index=1] {finalplottxt/h7.38kk.txt};
        \addplot+[dashed,thick]
        table[x index=0, y index=1] {finalplottxt/h6.s38kk.txt};
        \addplot+[dotted,thick]
        table[x index=0, y index=1] {finalplottxt/h7.s38kk.txt};
        \addplot+[solid]
        table[x index=0, y index=1] {finalplottxt/h8.s38kk.txt};
      \end{semilogyaxis}
    \end{tikzpicture}
  }
  \hfill
  \subfloat[$1$ GPU]{
    \label{fig:1gpu}
    \begin{tikzpicture}[font=\footnotesize]
      \begin{semilogyaxis}[
        height=.3\textheight,
        width=.5\textwidth,
        legend style={at={(.91,.96)},font=\tiny},
        xlabel=Accuracy $(Acc)$,
        legend columns=2,
        minor y tick num=3,
        grid=both,
        ymin=0,
        xtick=data,
        ymax=1200,
        ymin=9,
        xmin=1.5,
        xmax=7.5,
        ]
        \addplot+[dashed,thick]
        table[x index=0, y index=2] {finalplottxt/h6.38kk.txt};
        \addlegendentry{$h=6$, uniform};
        \addplot+[dotted,thick]
        table[x index=0, y index=2] {finalplottxt/h7.38kk.txt};
        \addlegendentry{$h=7$, uniform};
        \addplot+[dashed,thick]
        table[x index=0, y index=2] {finalplottxt/h6.s38kk.txt};
        \addlegendentry{$h=6$, non-uniform};
        \addplot+[dotted,thick]
        table[x index=0, y index=2] {finalplottxt/h7.s38kk.txt};
        \addlegendentry{$h=7$, non-uniform};
        \addplot+[solid]
        table[x index=0, y index=2] {finalplottxt/h8.s38kk.txt};
        \addlegendentry{$h=8$, non-uniform};
      \end{semilogyaxis}
    \end{tikzpicture}
  }
  \hfill
  \subfloat[$2$ GPUs]{
    \label{fig:2gpu}
    \begin{tikzpicture}[font=\footnotesize]
      \begin{semilogyaxis}[
        height=.3\textheight,
        width=.5\textwidth,
        ylabel=Time $(\unit{s})$,
        xlabel=Accuracy $(Acc)$,
        xtick=data,
        minor y tick num=3,
        grid=both,
        ymin=9,
        ymax=1200,
        xmin=1.5,
        xmax=7.5,
        ]
        \addplot+[dashed,thick]
        table[x index=0, y index=3] {finalplottxt/h6.38kk.txt};
        \addplot+[dotted,thick]
        table[x index=0, y index=3] {finalplottxt/h7.38kk.txt};
        \addplot+[dashed,thick]
        table[x index=0, y index=3] {finalplottxt/h6.s38kk.txt};
        \addplot+[dotted,thick]
        table[x index=0, y index=3] {finalplottxt/h7.s38kk.txt};
        \addplot+[solid]
        table[x index=0, y index=3] {finalplottxt/h8.s38kk.txt};
      \end{semilogyaxis}
    \end{tikzpicture}
  }
  \hfill
  \subfloat[$3$ GPUs]{
    \label{fig:3gpu}
    \begin{tikzpicture}[font=\footnotesize]
      \begin{semilogyaxis}[
        height=.3\textheight,
        width=.5\textwidth,
        xlabel=Accuracy $(Acc)$,
        minor y tick num=3,
        xtick=data,
        grid=both,
        ymin=9,
        ymax=1200,
        xmin=1.5,
        xmax=7.5,
        ]
        \addplot+[dashed,thick]
        table[x index=0, y index=4] {finalplottxt/h6.38kk.txt};
        \addplot+[dotted,thick]
        table[x index=0, y index=4] {finalplottxt/h7.38kk.txt};
        \addplot+[dashed,thick]
        table[x index=0, y index=4] {finalplottxt/h6.s38kk.txt};
        \addplot+[dotted,thick]
        table[x index=0, y index=4] {finalplottxt/h7.s38kk.txt};
        \addplot+[solid]
        table[x index=0, y index=4] {finalplottxt/h8.s38kk.txt};
      \end{semilogyaxis}
    \end{tikzpicture}
  }
  \caption{Comparison of timing results (logarithmic scale) for uniform and
    non-uniform distribution of $N=38\cdot10^6$ particles on $0,1,2$ or $3$
    GPUs and $n_g=1000$}
  \label{fig:times_acc}
\end{figure*}
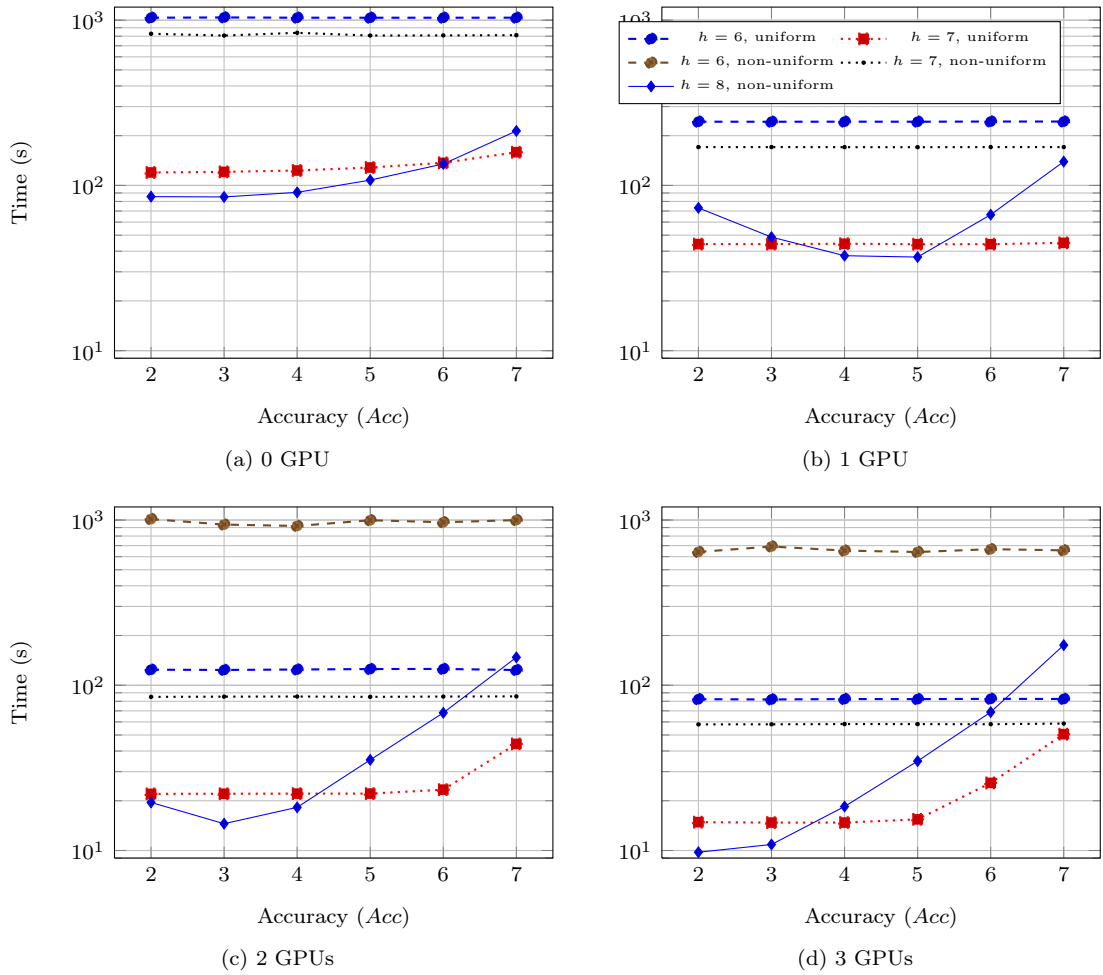

In Fig.~\ref{fig:times_acc} we compare absolute timing results corresponding
to the flop-rates from Fig.~\ref{fig:uniformflops} and
Fig.~\ref{fig:nonuniformflops}. We omit results for the uniform case with
$h=8$. In Fig.~\ref{fig:0gpu} and \ref{fig:1gpu} the times for the non-uniform
case and $h=6$ (dominated by near-field) are outside of the presentable
range. All figures show that in the uniform case a octree height of $h=7$
leads to the shortest computation times. In the non-uniform case for
accuracies $Acc<6$ an octree $h=8$ otherwise $h=7$ provides the best setting.

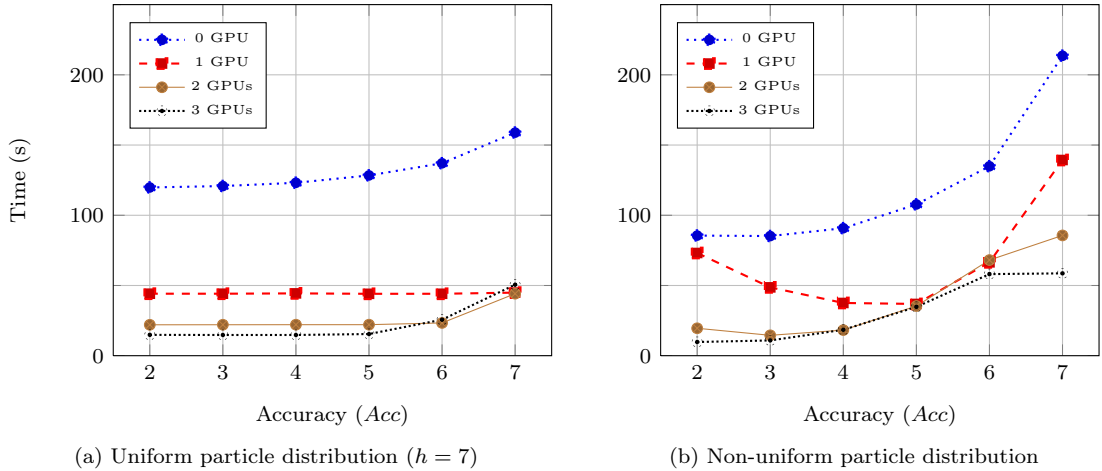
\begin{figure*}[htbp]
  \centering
  \subfloat[Uniform particle distribution $(h=7)$]{
    \label{fig:best_uniform}
    \begin{tikzpicture}[font=\footnotesize]
      \begin{axis}[
        height=.3\textheight,
        width=.5\textwidth,
        ylabel=Time $(\unit{s})$,
        xlabel=Accuracy $(Acc)$,
        minor y tick num=1,
        grid=both,
        legend style={at={(.35,.95)},font=\tiny},
        xtick=data,
        ymin=0,
        ymax=250,
        xmin=1.5,
        xmax=7.5,
        ]
        \addplot+[blue,dotted,thick]
        table[x index=0, y index=1] {finalplottxt/h7.38kk.txt};
        \addlegendentry{$0$ GPU};
        \addplot+[red,dashed,thick]
        table[x index=0, y index=2] {finalplottxt/h7.38kk.txt};
        \addlegendentry{$1$ GPU};
        \addplot+[brown,solid]
        table[x index=0, y index=3] {finalplottxt/h7.38kk.txt};
        \addlegendentry{$2$ GPUs};
        \addplot+[black,densely dotted,thick]
        table[x index=0, y index=4] {finalplottxt/h7.38kk.txt};
        \addlegendentry{$3$ GPUs};
      \end{axis}
    \end{tikzpicture}
  }
  \hfill
  \subfloat[Non-uniform particle distribution]{
    \label{fig:best_nonuniform}
    \begin{tikzpicture}[font=\footnotesize]
      \begin{axis}[
        height=.3\textheight,
        width=.5\textwidth,
        legend style={at={(.35,.95)},font=\tiny},
        xlabel=Accuracy $(Acc)$,
        minor y tick num=1,
        grid=both,
        ymin=0,
        xtick=data,
        ymax=250,
        xmin=1.5,
        xmax=7.5,
        ]
        \addplot+[blue,dotted,thick]
        table[x index=0, y index=1] {finalplottxt/minimal_timings.txt};
        \addlegendentry{$0$ GPU};
        \addplot+[red,dashed,thick]
        table[x index=0, y index=2] {finalplottxt/minimal_timings.txt};
        \addlegendentry{$1$ GPU};
        \addplot+[brown,solid]
        table[x index=0, y index=3] {finalplottxt/minimal_timings.txt};
        \addlegendentry{$2$ GPUs};
        \addplot+[black,densely dotted,thick]
        table[x index=0, y index=4] {finalplottxt/minimal_timings.txt};
        \addlegendentry{$3$ GPUs};
      \end{axis}
    \end{tikzpicture}
  }
  \caption{Minimal timing over all $h$ from Fig.~\ref{fig:times_acc}}
  \label{fig:best_times}
\end{figure*}

Figure~\ref{fig:best_times} addresses the scaling behavior of our dynamically
scheduled FMM implementation in the case of heterogeneous architectures.  We
take the results from Figure~\ref{fig:times_acc}, choose the minimal timings
for each $h$ and plot them over $Acc$. In the uniform case
(Figure~\ref{fig:best_uniform}) they correspond to $h=7$. In the non-uniform
case (Figure~\ref{fig:best_nonuniform}) and with $0$ and $1$ GPU they
correspond to $h=7$, too. However, with $2,3$ GPUs and up an accuracy
$Acc=5,6$ an octree height $h=8$ and for higher accuracies $h=7$ leads to
minimal timings. As expected, we get excellent scaling for low accuracies
($Acc<5$) where the near-field is dominating (we have an optimized GPU
implementation of the P2P kernel). For example, let us look at $Acc=3$:
without GPU the computation takes $\unit[120.8]{s}$, with $1,2,3$ GPUs it
takes $\unit[44.1]{s}$, $\unit[22.1]{s}$ and $\unit[14.7]{s}$,
respectively. The scaling for high accuracies can be improved by implementing
optimized far-field kernels (P2M, M2M, M2L, L2L and L2P).

A comment on the $1$ GPU case in Figure~\ref{fig:best_nonuniform}: the timings
for $Acc=2,3$ are not as expected. They should be the same as for $Acc=4,5$,
because the near-field is dominating up to $Acc\le5$. In our understanding the
reason for that is the fact that the \emph{Eager} scheduler (see Section
\ref{sec:heterogeneous}) has problems to schedule very small tasks
correctly. However, by using the \emph{Heft} scheduler we were able to
reproduce the expected results.


\section{Conclusion}
\label{sec:conclusion}

We have proposed an original FMM implementation over a runtime system.
Thanks to a breakdown of the computational work
(Section~\ref{sec:quantitative_breakdown}), we have shown that FMM
presents excellent properties for being efficiently pipelined. We have
succeeded to design a task flow that can exploit these properties and
achieve a very high parallel efficiency up to 160 CPU cores on a SGI
Altix UV 100 machine
(Section~\ref{sec:homogeneous-scalability}). Because almost the entire
work load is shared by P2P and M2L tasks
(Section~\ref{sec:quantitative_breakdown}), we have implemented a GPU version
of both these kernels. We have shown that when the near-field is
dominant (Section~\ref{sec:heterogeneous-nearfield}) or comparable to
the far-field (Section~\ref{sec:heterogeneous-intermediatenearfield}),
we manage to consistently pipeline the task flow and achieve very well
balanced executions by deporting P2P to GPU on demand. Thanks to a
highly optimized P2P kernel, we have achieved very high
performance. For instance, potentials and forces for $38$ million
particles could be processed in $\unit[13.34]{s}$ on our 12 cores
Nehalem processor enhanced with 3 M2090 Fermi GPUs. When the far-field
is dominant, some M2L tasks also have to be deported to GPUs in order
to balance the load. We have shown that an efficient pipeline could
be performed even in
that case (although the impact on performance itself is less
impressive since we did not optimize the GPU M2L kernel).

The successive optimizations discussed in this study result in a
single code that gets highly pipelined and balanced across
architectures. Since we rely on the so-called black-box method, our
code furthermore has a broad range of
applications (Section~\ref{sec:black-box-fmm}).

We plan to apply this approach to clusters of heterogeneous nodes. One
possibility will consist of statically distributing the work load to
the different nodes of the cluster and explicitly handling the
inter-node communications. Alternatively, we might also let the
runtime handle these communications thanks to the task flow. We also
plan to design optimized GPU kernels for the six FMM tasks. Indeed, if
all GPU kernels are provided to the runtime, it can limit data
movement by processing on individual GPUs connected subparts of the task
flow. It will be interesting to assess whether such an approach allows
for a better strong scaling.

\section*{Acknowledgment}

The authors would like to thank Raymond Namyst and Samuel Thibault for
their advice on performance optimization with StarPU.
Experiments presented in this paper were carried out using the PLAFRIM experimental testbed, being developed under the Inria PlaFRIM development action with support from LABRI and IMB and other entities: Conseil Régional d'Aquitaine, FeDER, Université de Bordeaux and CNRS (see https://plafrim.bordeaux.inria.fr/).

\bibliographystyle{IEEEtran}
\bibliography{fmm}

\begin{thebibliography}{10}
\providecommand{\url}[1]{#1}
\csname url@samestyle\endcsname
\providecommand{\newblock}{\relax}
\providecommand{\bibinfo}[2]{#2}
\providecommand{\BIBentrySTDinterwordspacing}{\spaceskip=0pt\relax}
\providecommand{\BIBentryALTinterwordstretchfactor}{4}
\providecommand{\BIBentryALTinterwordspacing}{\spaceskip=\fontdimen2\font plus
\BIBentryALTinterwordstretchfactor\fontdimen3\font minus
  \fontdimen4\font\relax}
\providecommand{\BIBforeignlanguage}[2]{{%
\expandafter\ifx\csname l@#1\endcsname\relax
\typeout{** WARNING: IEEEtran.bst: No hyphenation pattern has been}%
\typeout{** loaded for the language `#1'. Using the pattern for}%
\typeout{** the default language instead.}%
\else
\language=\csname l@#1\endcsname
\fi
#2}}
\providecommand{\BIBdecl}{\relax}
\BIBdecl

\bibitem{greengard:87}
\BIBentryALTinterwordspacing
L.~Greengard and V.~Rokhlin, ``A fast algorithm for particle simulations,''
  \emph{Journal of Computational Physics}, vol.~73, no.~2, pp. 325 -- 348, Dec.
  1987. [Online]. Available:
  \url{http://dx.doi.org/10.1016/0021-9991(87)90140-9}
\BIBentrySTDinterwordspacing

\bibitem{greengard97}
------, ``A new version of the fast multipole method for the laplace equation
  in three dimensions,'' \emph{Acta Numerica}, vol.~6, pp. 229--269, 1997.

\bibitem{Greengard199063}
L.~Greengard and W.~D. Gropp, ``A parallel version of the fast multipole
  method,'' \emph{Computers \& Mathematics with Applications}, vol.~20, no.~7,
  pp. 63 -- 71, 1990.

\bibitem{Chandramowlishwaran2010a}
A.~Chandramowlishwaran, S.~Williams, L.~Oliker, a.~G.~B. Ilya~Lashuk, and
  R.~Vuduc, ``Optimizing and tuning the fast multipole method for
  state-of-the-art multicore architectures,'' in \emph{Proceedings of the 2010
  IEEE conference of IPDPS}, 2010, pp. 1--15.

\bibitem{Cruz2011}
F.~A. Cruz, M.~G. Knepley, and L.~A. Barba, ``Petfmm—a dynamically
  load-balancing parallel fast multipole library,'' \emph{Int. J. Numer. Meth.
  Engng.}, vol.~85, no.~4, p. 403–428, 2011.

\bibitem{EricDarve2011}
E.~Darve, C.~Cecka, and T.~Takahashi, ``The fast multipole method on parallel
  clusters, multicore processors, and graphics processing units,''
  \emph{Comptes Rendus Mécanique}, vol. 339, no. 2-3, pp. 185--193, 2011.

\bibitem{Yokota20092066}
\BIBentryALTinterwordspacing
R.~Yokota, T.~Narumi, R.~Sakamaki, S.~Kameoka, S.~Obi, and K.~Yasuoka, ``Fast
  multipole methods on a cluster of gpus for the meshless simulation of
  turbulence,'' \emph{Computer Physics Communications}, vol. 180, no.~11, pp.
  2066 -- 2078, 2009. [Online]. Available:
  \url{http://www.sciencedirect.com/science/article/pii/S0010465509001891}
\BIBentrySTDinterwordspacing

\bibitem{Gumerov2008}
\BIBentryALTinterwordspacing
N.~A. Gumerov and R.~Duraiswami, ``Fast multipole methods on graphics
  processors,'' \emph{Journal of Computational Physics}, vol. 227, no.~18, pp.
  8290 -- 8313, 2008. [Online]. Available:
  \url{http://www.sciencedirect.com/science/article/pii/S0021999108002921}
\BIBentrySTDinterwordspacing

\bibitem{42tflops}
\BIBentryALTinterwordspacing
T.~Hamada, T.~Narumi, R.~Yokota, K.~Yasuoka, K.~Nitadori, and M.~Taiji,
  \emph{42 TFlops hierarchical N-body simulations on GPUs with applications in
  both astrophysics and turbulence}.\hskip 1em plus 0.5em minus 0.4em\relax
  ACM, 2009, pp. 62:1---62:12. [Online]. Available:
  \url{http://doi.acm.org/10.1145/1654059.1654123}
\BIBentrySTDinterwordspacing

\bibitem{takahashi:2012}
\BIBentryALTinterwordspacing
T.~Takahashi, C.~Cecka, W.~Fong, and E.~Darve, ``Optimizing the
  multipole-to-local operator in the fast multipole method for graphical
  processing units,'' \emph{International Journal for Numerical Methods in
  Engineering}, vol.~89, no.~1, pp. 105--133, 2012. [Online]. Available:
  \url{http://dx.doi.org/10.1002/nme.3240}
\BIBentrySTDinterwordspacing

\bibitem{Hu:2011}
\BIBentryALTinterwordspacing
Q.~Hu, N.~A. Gumerov, and R.~Duraiswami, ``Scalable fast multipole methods on
  distributed heterogeneous architectures,'' in \emph{Proceedings of 2011
  International Conference for High Performance Computing, Networking, Storage
  and Analysis}, ser. SC '11.\hskip 1em plus 0.5em minus 0.4em\relax New York,
  NY, USA: ACM, 2011, pp. 36:1--36:12. [Online]. Available:
  \url{http://doi.acm.org/10.1145/2063384.2063432}
\BIBentrySTDinterwordspacing

\bibitem{Lashuki2009}
\BIBentryALTinterwordspacing
I.~Lashuk, C.~Aparna, H.~Langston, T.-A. Nguyen, R.~Sampath, A.~Shringarpure,
  R.~Vuduc, l.~Ying, D.~Zorin, and G.~Biros, ``A massively parallel adaptive
  fast-multipole method on heterogeneous architectures,'' in \emph{Proceedings
  of the 2009 ACM/IEEE conference on Supercomputing}, 2009, p. 1–11.
  [Online]. Available: \url{http://www.cc.gatech.edu/~gbiros/}
\BIBentrySTDinterwordspacing

\bibitem{faster}
\BIBentryALTinterwordspacing
E.~Agullo, C.~Augonnet, J.~Dongarra, H.~Ltaief, R.~Namyst, S.~Thibault, and
  S.~Tomov, ``\BIBforeignlanguage{Anglais}{{Faster, Cheaper, Better -- a
  Hybridization Methodology to Develop Linear Algebra Software for GPUs}},'' in
  \emph{\BIBforeignlanguage{Anglais}{{GPU Computing Gems}}}, W.~mei W.~Hwu,
  Ed.\hskip 1em plus 0.5em minus 0.4em\relax Morgan Kaufmann, Sep. 2010,
  vol.~2. [Online]. Available: \url{http://hal.inria.fr/inria-00547847}
\BIBentrySTDinterwordspacing

\bibitem{LU}
\BIBentryALTinterwordspacing
E.~Agullo, C.~Augonnet, J.~Dongarra, M.~Faverge, J.~Langou, H.~Ltaief, and
  S.~Tomov, ``{LU} factorization for accelerator-based systems,'' in \emph{The
  9th {IEEE}/{ACS} International Conference on Computer Systems and
  Applications, {AICCSA} 2011, Sharm El-Sheikh, Egypt, December 27-30, 2011},
  H.~J. Siegel and A.~El-Kadi, Eds.\hskip 1em plus 0.5em minus 0.4em\relax
  IEEE, 2011, pp. 217--224. [Online]. Available:
  \url{http://ieeexplore.ieee.org/xpl/mostRecentIssue.jsp?punumber=6122397}
\BIBentrySTDinterwordspacing

\bibitem{QR}
\BIBentryALTinterwordspacing
E.~Agullo, C.~Augonnet, J.~Dongarra, M.~Faverge, H.~Ltaief, S.~Thibault, and
  S.~Tomov, ``{QR} factorization on a multicore node enhanced with multiple
  {GPU} accelerators,'' in \emph{IPDPS}.\hskip 1em plus 0.5em minus 0.4em\relax
  IEEE, 2011, pp. 932--943. [Online]. Available:
  \url{http://ieeexplore.ieee.org/xpl/mostRecentIssue.jsp?punumber=6011824}
\BIBentrySTDinterwordspacing

\bibitem{Quintana-Orti}
G.~Quintana-Ort{\'\i}, F.~D. Igual, E.~S. Quintana-Ort{\'\i}, and R.~A. van~de
  Geijn, ``Solving dense linear systems on platforms with multiple hardware
  accelerators,'' \emph{ACM SIG{\-}PLAN Notices}, vol.~44, no.~4, pp. 121--130,
  Apr. 2009.

\bibitem{quintana-orti2008scheduling}
G.~Quintana-Ort{\'{\i}}, E.~S. Quintana-Ort{\'{\i}}, E.~Chan, F.~G.~V. Zee, and
  R.~A. van~de Geijn, ``Scheduling of {QR} factorization algorithms on {SMP}
  and multi-core architectures,'' in \emph{Proceedings of PDP'08}, 2008,
  {FLAME} Working Note \#24.

\bibitem{tileqr}
A.~Buttari, J.~Langou, J.~Kurzak, and J.~Dongarra, ``{Parallel tiled QR
  factorization for multicore architectures},'' \emph{Concurrency and
  Computation: Practice and Experience}, vol.~20, no.~13, pp. 1573--1590, 2008.

\bibitem{SDLA}
\BIBentryALTinterwordspacing
J.~Kurzak, H.~Ltaief, J.~Dongarra, and R.~M. Badia, ``Scheduling dense linear
  algebra operations on multicore processors,'' \emph{Concurrency and
  Computation: Practice and Experience}, vol.~22, no.~1, pp. 15--44, 2010.
  [Online]. Available: \url{http://dx.doi.org/10.1002/cpe.1467}
\BIBentrySTDinterwordspacing

\bibitem{BosilcaBDFHHKLLLLYD11}
\BIBentryALTinterwordspacing
G.~Bosilca, A.~Bouteiller, A.~Danalis, M.~Faverge, A.~Haidar, T.~H{\'e}rault,
  J.~Kurzak, J.~Langou, P.~Lemarinier, H.~Ltaief, P.~Luszczek, A.~YarKhan, and
  J.~Dongarra, ``Flexible development of dense linear algebra algorithms on
  massively parallel architectures with {DPLASMA},'' in \emph{IPDPS
  Workshops}.\hskip 1em plus 0.5em minus 0.4em\relax IEEE, 2011, pp.
  1432--1441. [Online]. Available:
  \url{http://ieeexplore.ieee.org/xpl/mostRecentIssue.jsp?punumber=6008655}
\BIBentrySTDinterwordspacing

\bibitem{PLASMA}
``{PLASMA} users' guide, parallel linear algebra software for multicore
  architectures, version 2.0,'' http://icl.cs.utk.edu/plasma, University of
  Tennessee, November 2009.

\bibitem{MAGMA}
``{MAGMA} users’ guide, version 0.2,'' http://icl.cs.utk.edu/magma,
  University of Tennessee, November 2009.

\bibitem{FLAME}
F.~G. {Van Zee}, E.~Chan, R.~A. van~de Geijn, E.~S. Quintana-Orti, and
  G.~Quintana-Orti, ``The {{\tt libflame}} library for dense matrix
  computations,'' \emph{Computing in Science and Engineering}, vol.~11, no.~6,
  pp. 56--63, Nov.\slash Dec. 2009.

\bibitem{ipdps2011}
E.~Agullo, C.~Augonnet, J.~Dongarra, M.~Faverge, H.~Ltaief, S.~Thibault, and
  S.~Tomov, ``{QR} factorization on a multicore node enhanced with multiple
  {GPU} accelerators,'' \emph{Accepted to the 25th {IEEE} International
  Parallel and Distributed Processing Symposium (IPDPS 2011)}, 2011.

\bibitem{ltaeif-yokota-2012}
H.~Ltaief and R.~Yokota, ``Data-driven execution of fast multipole methods,''
  \emph{CoRR}, vol. abs/1203.0889, 2012.

\bibitem{bordage-2012}
C.~Bordage, 2012, personal communication.

\bibitem{AugThiNamWac10CCPE}
C.~Augonnet, S.~Thibault, R.~Namyst, and P.-A. Wacrenier, ``{StarPU: A Unified
  Platform for Task Scheduling on Heterogeneous Multicore Architectures},''
  \emph{Concurrency and Computation: Practice and Experience, Euro-Par 2009
  best papers issue}, 2010.

\bibitem{Fong2009}
\BIBentryALTinterwordspacing
W.~Fong and E.~Darve, ``The black-box fast multipole method,'' \emph{Journal of
  Computational Physics}, vol. 228, no.~23, pp. 8712 -- 8725, 2009. [Online].
  Available:
  \url{http://www.sciencedirect.com/science/article/pii/S0021999109004665}
\BIBentrySTDinterwordspacing

\bibitem{Ying:2004}
\BIBentryALTinterwordspacing
L.~Ying, G.~Biros, and D.~Zorin, ``A kernel-independent adaptive fast multipole
  algorithm in two and three dimensions,'' \emph{J. Comput. Phys.}, vol. 196,
  no.~2, pp. 591--626, May 2004. [Online]. Available:
  \url{http://dx.doi.org/10.1016/j.jcp.2003.11.021}
\BIBentrySTDinterwordspacing

\bibitem{Messner2012}
\BIBentryALTinterwordspacing
M.~Messner, M.~Schanz, and E.~Darve, ``Fast directional multilevel summation
  for oscillatory kernels based on chebyshev interpolation,'' \emph{Journal of
  Computational Physics}, vol. 231, no.~4, pp. 1175 -- 1196, 2012. [Online].
  Available:
  \url{http://www.sciencedirect.com/science/article/pii/S0021999111005705}
\BIBentrySTDinterwordspacing

\bibitem{GPUSs09Europar}
E.~Ayguad\'{e}, R.~M. Badia, F.~D. Igual, J.~Labarta, R.~Mayo, and E.~S.
  Quintana-Ort\'{\i}, ``{An Extension of the StarSs Programming Model for
  Platforms with Multiple GPUs},'' in \emph{{Proceedings of the 15th
  International Euro-Par Conference on Parallel Processing}}.\hskip 1em plus
  0.5em minus 0.4em\relax Berlin, Heidelberg: Springer-Verlag, 2009, pp.
  851--862.

\bibitem{Bosilca201237}
\BIBentryALTinterwordspacing
G.~Bosilca, A.~Bouteiller, A.~Danalis, T.~Herault, P.~Lemarinier, and
  J.~Dongarra, ``Dague: A generic distributed dag engine for high performance
  computing,'' \emph{Parallel Computing}, vol.~38, no. 1–2, pp. 37 -- 51,
  2012, <ce:title>Extensions for Next-Generation Parallel Programming
  Models</ce:title>. [Online]. Available:
  \url{http://www.sciencedirect.com/science/article/pii/S0167819111001347}
\BIBentrySTDinterwordspacing

\bibitem{1383447}
G.~F. Diamos and S.~Yalamanchili, ``Harmony: an execution model and runtime for
  heterogeneous many core systems,'' in \emph{HPDC '08: Proceedings of the 17th
  international symposium on High performance distributed computing}.\hskip 1em
  plus 0.5em minus 0.4em\relax New York, NY, USA: ACM, 2008, pp. 197--200.

\bibitem{SEQUOIA_SC06}
K.~Fatahalian, T.~Knight, M.~Houston, M.~Erez, D.~Horn, L.~Leem, J.~Park,
  M.~Ren, A.~Aiken, W.~Dally, and P.~Hanrahan, ``Sequoia: Programming the
  memory hierarchy,'' in \emph{{ACM/IEEE} SC'06 {C}onference}, 2006.

\bibitem{2009ChaNGaGPU}
P.~Jetley, L.~Wesolowski, F.~Gioachin, L.~V. Kale, and T.~R. Quinn, ``Scaling
  hierarchical {N}-body simulations on {GPU} clusters,'' in \emph{SC'10 USB
  Key}.\hskip 1em plus 0.5em minus 0.4em\relax New Orleans, LA, USA: ACM/IEEE,
  Nov. 2010.

\bibitem{Warren:1993}
\BIBentryALTinterwordspacing
M.~S. Warren and J.~K. Salmon, ``A parallel hashed oct-tree n-body algorithm,''
  in \emph{Proceedings of the 1993 ACM/IEEE conference on Supercomputing}, ser.
  Supercomputing '93.\hskip 1em plus 0.5em minus 0.4em\relax New York, NY, USA:
  ACM, 1993, pp. 12--21. [Online]. Available:
  \url{http://doi.acm.org/10.1145/169627.169640}
\BIBentrySTDinterwordspacing

\bibitem{AugThiNamWac11CCPE}
\BIBentryALTinterwordspacing
C.~Augonnet, S.~Thibault, R.~Namyst, and P.-A. Wacrenier, ``{StarPU: A Unified
  Platform for Task Scheduling on Heterogeneous Multicore Architectures},''
  \emph{Concurrency and Computation: Practice and Experience, Special Issue:
  Euro-Par 2009}, vol.~23, pp. 187--198, Feb. 2011. [Online]. Available:
  \url{http://hal.inria.fr/inria-00550877}
\BIBentrySTDinterwordspacing

\end{thebibliography}

\end{document}